\documentclass[preprint,nofootinbib,superscriptaddress]{revtex4-1}
\usepackage{graphicx}
\usepackage{dcolumn}
\usepackage{bm}
\usepackage[normalem]{ulem}
\usepackage{amsmath}
\usepackage{amsfonts} 
\usepackage{latexsym}
\usepackage{bbm}
\usepackage{color}
\usepackage{xcolor}
\usepackage{booktabs} 
\usepackage{amssymb}
\usepackage{amsthm}
\usepackage{caption3}
\usepackage{appendix}
\usepackage{mathtools}
\usepackage{cases}
\usepackage{subfigure}
\usepackage{multirow}
\usepackage{comment}

\usepackage{mathrsfs}
\usepackage{hyperref}
\hypersetup{
  setpagesize=false,
  bookmarksnumbered=true,
  bookmarksopen=true,
  colorlinks=true,
  linkcolor=blue,
  citecolor=blue,
}
\makeatletter

\newcommand{\Rmnum}[1]{\expandafter\@slowromancap\romannumeral #1@}
\makeatother

\begin{document}
\title{Model-Parameter Reconstruction of Electroweak Phase Transition with TianQin and LISA: Insights from the Dimension-Six Model}
\author{Aidi Yang}
\author{Chikako Idegawa}
\author{Fa Peng Huang}
\email{ Corresponding Author.
 huangfp8@sysu.edu.cn }

\affiliation{MOE Key Laboratory of TianQin Mission, TianQin Research Center for
Gravitational Physics \& School of Physics and Astronomy, Frontiers
Science Center for TianQin, Gravitational Wave Research Center of CNSA, 
Sun Yat-sen University (Zhuhai Campus), Zhuhai 519082, China}

\begin{abstract}
We investigate the capability of TianQin and LISA to reconstruct the model parameters in the Lagrangian of new physics scenarios that can generate an electroweak SFOPT.
Taking the dimension-six Higgs operator extension of the Standard Model as a representative scenario for a broad class of new physics models, we establish the mapping between the model parameter $\Lambda$ and the observable spectral features of the stochastic gravitational wave background. We begin by generating simulated data incorporating Time Delay Interferometry channel noise, astrophysical foregrounds, and signals from the dimension-six model. The data are then compressed and optimized, followed by geometric parameter inference using both Fisher matrix analysis and Bayesian nested sampling with \texttt{PolyChord}, which efficiently handles high-dimensional, multimodal posterior distributions. Finally, machine-learning techniques are employed to achieve precise reconstruction of the model parameter $\Lambda$.
For benchmark points producing strong signals, parameter reconstruction with both TianQin and LISA yields relative uncertainties of approximately $20$--$30\%$ in the signal amplitude and sub-percent precision in the model parameter $\Lambda$. The sub-percent precision reflects the statistical reconstruction capability of the detectors in an idealized setting: it incorporates the machine-learning inference uncertainty and is established at a fixed bubble wall velocity, while theoretical uncertainties in the effective potential calculation are not included.
\end{abstract}

\maketitle

\section{Introduction}
Since the first direct detection of gravitational waves (GWs) from a binary black hole merger by the LIGO–Virgo Collaboration in 2015~\cite{LIGOScientific:2016aoc}, GW cosmology has emerged as a powerful tool for probing the fundamental problems in particle cosmology. The upcoming space-based interferometers  LISA~\cite{LISACosmologyWorkingGroup:2022jok}, TianQin~\cite{TianQin:2015yph,Luo:2025ewp}, and Taiji~\cite{Hu:2017mde}  are expected to deliver high-precision measurements in the milli-Hertz frequency band, opening new avenues for exploring cosmological and particle physics phenomena. Beyond conventional astrophysical sources, GWs offer a unique observational window into fundamental physics, enabling the reconstruction of model parameters in the Lagrangian—particularly those associated with phase transitions in the early universe~\cite{Caprini:2024hue,Blanco-Pillado:2024aca,LISACosmologyWorkingGroup:2024hsc,Boileau:2022ter,Giese:2021dnw,Gowling:2022pzb,Liang:2024tgn, Huang:2025uer,LIGOScientific:2017bnn}. 

The motivation for using GW observations to probe beyond the Standard Model (BSM) physics arises from fundamental limitations of the Standard Model (SM). While the SM has achieved remarkable success, it cannot explain dark matter, dark energy, or the observed matter-antimatter asymmetry in the universe. These shortcomings point to new physics operating at higher energy scales or during the early universe. In particular, for the observed $125~\mathrm{GeV}$ Higgs boson mass, the SM predicts that the electroweak phase transition (EWPT) is a crossover~\cite{Kajantie:1996mn, Rummukainen:1998as, Csikor:1998eu}, which cannot generate detectable GW. This motivates the study of different types of BSM scenarios, which aim to explain the dark matter or the observed matter-antimatter asymmetry. Among these BSM scenarios, various new physics models can modify the Higgs potential and make EWPT a strong first-order phase transition (SFOPT). A well-motivated approach is to extend the SM with dimension-six Higgs operators in an effective field theory framework~\cite{Zhang:1992fs,Grojean:2004xa,Huang:2015izx,Huang:2016odd,deVries:2017ncy,Delaunay:2007wb,Cao:2017oez,Huang:2025cqi}. Such operators naturally arise by integrating out heavy degrees of freedom from various ultraviolet (UV)-complete theories—including singlet extensions,
composite Higgs models, and two-Higgs-doublet models~\cite{Cao:2017oez}. Crucially, they can convert the EWPT into a SFOPT, making it an observable source of GWs for space-based detectors. A key advantage of this framework is its parametric simplicity: the phase transition dynamics are controlled only by a single parameter, $\Lambda$, allowing GW observations to be directly translated into constraints on BSM physics while avoiding the complex parameter degeneracies inherent in multi-parameter models.

During a SFOPT, the universe transitions from a high-energy false vacuum to a true vacuum state through bubble nucleation.
Phase transition GWs can be produced from three primary mechanisms: bubble collisions~\cite{Caprini:2015zlo,Huber:2008hg}, sound waves (SWs) generated in the plasma~\cite{Hindmarsh:2016lnk,Konstandin:2017sat,Hindmarsh:2019phv,Jinno:2020eqg,Jinno:2022mie}, and the magnetohydrodynamic (MHD) turbulence~\cite{Caprini:2009yp,RoperPol:2019wvy}, which together produce a stochastic gravitational-wave background (SGWB). Recently, a new source of phase transition GWs has been discovered from heavy particles braking across bubble walls in
Refs.~\cite{Qiu:2025tmn,Ai:2025fqw}.
In a non-runaway regime, the contribution to the production of GW from the
bubble collision source is subdominant~\cite{Bodeker:2009qy}.  The MHD turbulence effects can contribute, particularly for strong phase transitions where nonlinear fluid dynamics become significant.  Recent numerical simulations suggest that the SW contribution remains the leading source for the range of phase transition strengths considered here~\cite{Giese:2021dnw, Boileau:2022ter, Gowling:2022pzb}.
Therefore, to be simple and consistent with previous studies~\cite{Giese:2021dnw, Boileau:2022ter, Gowling:2022pzb}, we focus on the SW mechanism in our analysis.

The peak frequency of GW signal from an electroweak SFOPT is expected to lie within the milli-Hertz frequency band, corresponding to the electroweak energy scale. This frequency range represents a primary detection target for the future space-based GW observatories TianQin~\cite{TianQin:2015yph} and LISA~\cite{LISACosmologyWorkingGroup:2022jok}. Consequently, the detection of GWs from an electroweak SFOPT would constitute direct evidence for BSM physics.  
However, the detection of a GW signal alone does not guarantee a comprehensive understanding of the underlying physics. Accurate reconstruction of model parameters from observed GW signals is crucial for probing BSM scenarios. This reconstruction serves as a vital bridge between GW observations and fundamental theoretical frameworks, offering insights that complement those obtained from particle physics experiments.

\begin{figure}[h!]
    \centering
\includegraphics[width=1\linewidth]{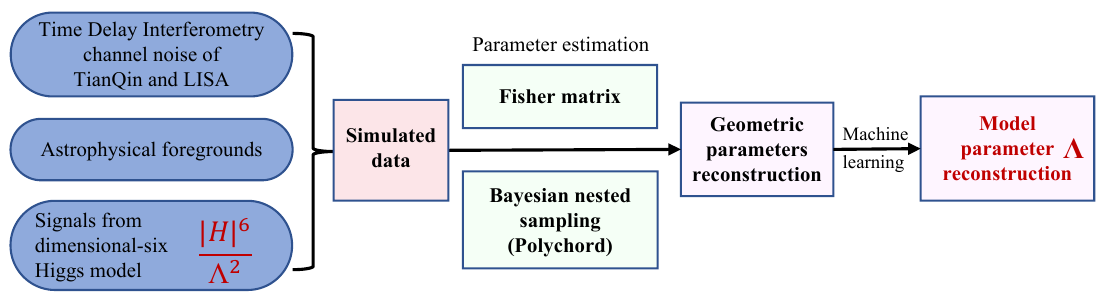}
    \caption{Schematic overview of the parameter reconstruction pipeline used to extract the model parameter $\Lambda$ with TianQin and LISA.}
    \label{fig:flow}
\end{figure}

Therefore, the central objective of this study is to assess whether space-based detectors such as TianQin or LISA can reliably reconstruct the model parameter $\Lambda$ from a detected GW signal, thereby placing meaningful constraints on new physics.
Following the pipeline shown in Fig.~\ref{fig:flow},
we compare the predicted SGWB spectrum from the dimension-six model with the sensitivity curves of both detectors and evaluate their ability to reconstruct model parameters. We reconstruct the geometric parameters using Fisher matrix  analysis~\cite{Gowling:2021gcy} and nested sampling with \texttt{PolyChord}~\cite{handley2015polychord,Handley:2015fda}, and subsequently use machine-learning to map these geometric parameters to the model parameter~\cite{Breiman:2001hzm,hastie2009elements,sivia2006data}.

Building on the reconstruction methods of Refs.~\cite{Giese:2021dnw,Boileau:2022ter,Gowling:2022pzb}, which reconstruct template parameters of the GW spectrum, we 
establish a direct mapping from GW observables to the fundamental model parameter 
$\Lambda$ in the dimension-six model.
For the first time, we perform a complete parameter reconstruction pipeline for TianQin, and perform the quantitative comparison between TianQin and LISA's reconstruction capabilities. 
To offer more model-independent
insights into the GW detection of the Higgs potential and the corresponding strong first-order EWPT models,
we use the dimension-six Higgs operator framework, which naturally arises within effective field theory and offers a more model-independent description of BSM physics. We adopt the SW template developed in Ref.~\cite{Caprini:2024hue} based on Ref.~\cite{Jinno:2022mie}. Note that the SW template continues to be an active area of research~\cite{RoperPol:2023dzg,Sharma:2023mao}.

This paper is organized as follows: Sec.~\ref{sec:model} provides an introduction to the EWPT in the dimension-six Higgs model, including the construction of the finite-temperature effective potential and the SGWB templates generated by the SW mechanism. Sec.~\ref{sec:map} clarifies the mapping from GW geometric parameters to model parameters and analyzes parameter degeneracies. Sec.~\ref{sec:noise} introduces the noise model and the parameter reconstruction pipeline for TianQin, including the application of the Fisher matrix and \texttt{PolyChord}. Sec.~\ref{sec:res_tq} presents the main reconstruction results, quantitatively evaluating the reconstruction capabilities of TianQin for the dimension-six model parameter. Sec.~\ref{sec:res_lisa} presents the main reconstruction results of LISA and compares the performance differences between the two detectors. Finally, conclusions and discussions are given in Sec.~\ref{sec:sum}.

\section{Electroweak phase transition gravitational wave in dimension-six model}\label{sec:model}

\subsection{Electroweak phase transition in the dimension-six model}
The SM predicts that the EWPT is a crossover and therefore incapable of generating an observable GW signal~\cite{Kajantie:1996mn,Rummukainen:1998as,Csikor:1998eu}. To address fundamental questions such as the origin of dark matter and the baryon asymmetry of the observable universe, it is often necessary to invoke physics BSM. Many such extensions allow for a SFOPT~\cite{Zhang:1992fs,Grojean:2004xa,Huang:2015izx,Huang:2016odd,deVries:2017ncy,Delaunay:2007wb, Cao:2017oez,Huang:2025cqi}. To ensure that our parameter reconstruction analysis remains largely independent of specific new physics scenarios, we adopt the framework of the Standard Model Effective Field Theory (SMEFT). In particular, we parameterize new physics effects in a model-independent manner by introducing a dimension-six operator $|H|^6$ in the Higgs potential, with the corresponding Wilson coefficient denoted by $c_6$~\cite{Huang:2015izx,Ellis:2018mja,Wang:2020jrd,Cao:2017oez,Wang:2020nzm,Wang:2020zlf,Wang:2021dwl, Athron:2023rfq}.
At dimension-six order, the effective Lagrangian includes the following correction to the Higgs potential~\cite{Cao:2017oez}
\begin{equation}
\mathcal{L} \supset m^2|H|^2-\lambda|H|^4-c_6|H|^6+\sum_i c_i \mathcal{O}_i.
\end{equation}
Here, $\mathcal{O}_i$ denotes other dimension-six operators in SMEFT, and $c_i$ are their corresponding Wilson coefficients. One can expand the Higgs doublet as
\begin{equation}
H=\frac{1}{\sqrt{2}}\binom{\chi_1+i \chi_2}{h_0+v+i \chi_3}.
\end{equation}
Here, $v = 246~\mathrm{GeV}$ denotes the vacuum expectation value at zero temperature. 
To avoid confusion with the dimensionless Hubble parameter, we denote the Higgs boson as $h_0$ throughout this work.
For simplicity, we choose 
$c_6=\frac{1}{\Lambda^2}$.
In the framework of SMEFT, the tree-level Higgs potential can be written as
\begin{equation}
V_{\text {tree}}(h_0)=-\frac{m^2}{2} h_0^2+\frac{\lambda}{4} h_0^4+\frac{1}{8} \frac{h_0^6}{\Lambda^2} .
\end{equation}

The finite-temperature effective potential at one-loop level is composed of three parts~\cite{Quiros:1999jp, Dolan:1973qd}
\begin{equation}
V_{\text{eff}}(h_0, T)=V_{\text {tree}}(h_0)+V_{\text{1-loop}}(h_0)+V_T(h_0, T),
\end{equation}
where $V_{\text {1-loop }}(h_0)$ is the one-loop quantum correction at zero temperature, $V_T(h_0, T)$ is the finite-temperature correction.
We use the on-shell scheme for the renormalization condition
\begin{equation}
V_{\text{eff}}^{\prime}(h_0=v)=0, \quad V_{\text{eff}}^{\prime \prime}(h_0=v)=m_h^2,
\end{equation}
where $m_h=125~\mathrm{GeV}$ from the observed data. Thus, one can obtain
\begin{equation}
m^2=\frac{m_h^2}{2}-\frac{3 v^4}{4 \Lambda^2}, \quad \lambda=\frac{m_h^2}{2 v^2}-\frac{3 v^2}{2 \Lambda^2} .
\end{equation}

Using the on-shell renormalization scheme, the one-loop corrections to the zero-temperature potential are given by~\cite{Delaunay:2007wb,Curtin:2014jma}
\begin{equation}
V_{\text{1-loop}}(h_0)=\sum_{i=h_0, \chi, W, Z, t} \frac{n_i}{64 \pi^2}\left[m_i^4\left(\log \frac{m_i^2}{m_{0 i}^2}-\frac{3}{2}\right)+2 m_i^2 m_{0 i}^2\right],
\end{equation}
where the field-dependent masses of the relevant particles are denoted by $m_i(h_0)$. The corresponding physical masses at zero temperature are defined as $m_{0i} \equiv m_i(v)$. The degrees of freedom for each particle are given by $n_{\{h_0, \chi, W, Z, t\}} = \{1, 3, 6, 3, -12\}$. The summation includes the Higgs boson $h_0$, the three Goldstone bosons $\chi_{1,2,3}$, the gauge bosons $W^{\pm}$ and $Z$, and the top quark $t$. The explicit forms of the field-dependent masses are as follows
\begin{equation}
\begin{aligned}
& m_h^2=-m^2+3 \lambda h_0^2+\frac{15}{4} \frac{h_0^4}{\Lambda^2}, \quad m_\chi^2=-m^2+\lambda h_0^2+\frac{3}{4} \frac{h_0^4}{\Lambda^2},\\
& m_W^2=\frac{g^2}{4} h_0^2, \quad m_Z^2=\frac{g^2+g^{\prime 2}}{4} h_0^2, \quad m_t^2=\frac{y_t^2}{2} h_0^2.
\end{aligned}
\end{equation}
Here, $g$ and $g^{\prime}$ denote the $SU(2)_L$ and $U(1)_Y$ gauge couplings, respectively. $y_t$ is the top quark Yukawa coupling. The finite-temperature correction can be written as
\begin{equation}
V_T(h_0, T)=\sum_{i=h_0, \chi, W, Z, \gamma} \frac{n_i T^4}{2 \pi^2} J_b\left(\frac{m_i^2}{T^2}\right)+\sum_{i=t} \frac{n_i T^4}{2 \pi^2} J_f\left(\frac{m_i^2}{T^2}\right),
\end{equation}
where the thermal contributions of bosons and fermions are evaluated using distinct thermal integral functions for each type $J_b$ and $J_f$
\begin{equation}
J_{b / f}\left(\frac{m_i^2}{T^2}\right)=\int_0^{\infty} d k k^2 \log \left[1 \mp \exp \left(-\sqrt{\frac{k^2+m_i^2}{T^2}}\right)\right].
\end{equation}
The one-loop thermal corrections discussed above suffer from infrared divergences. To properly account for this physical effect, it is necessary to include the contributions from ring diagrams, commonly referred to as the daisy resummation. This resummation is implemented through a thermal mass correction: for the longitudinal polarization modes of gauge bosons and scalar bosons, we adopt the Parwani scheme~\cite{Parwani:1991gq}, replacing the $m_i^2$ with $m_i^2 + \Pi_i(T)$ in the loop calculations. In our model, the thermal mass-squared corrections take the following form~\cite{Delaunay:2007wb,Carrington:1991hz}
\begin{equation}
\begin{aligned}
\Pi_{h_0, \chi}(T) & =\frac{T^2}{4 v^2}\left(m_h^2+2 m_W^2+m_Z^2+2 m_t^2\right)-\frac{3 T^2 v^2}{4 \Lambda^2}, \\
\Pi_W(T) & =\frac{22}{3} \frac{m_W^2}{v^2} T^2.
\end{aligned}
\end{equation}
The thermal masses of the $Z$ boson and the photon are obtained by diagonalizing the gauge boson mixing matrix
\begin{equation}
\left(\begin{array}{cc}\frac{1}{4} g^2 h_0^2+\frac{11}{6} g^2 T^2 & -\frac{1}{4} g^{\prime } g h_0^2 \\ -\frac{1}{4} g^{\prime } g h_0^2 & \frac{1}{4} g^{\prime 2} h_0^2+\frac{11}{6} g^{\prime 2} T^2\end{array}\right).
\end{equation}

The introduced dimension-six operator modifies the evolution of the Higgs potential at finite temperature. After including the one-loop thermal corrections, the conditions for a SFOPT are satisfied~\cite{Bodeker:2004ws}. This phase transition process can be characterized by the following phase transition parameters~\cite{Wang:2020jrd}
 \begin{itemize}
       \item The percolation temperature $T_p$: The temperature at which 34\% of the false vacuum has converted to the true vacuum.
	\item Phase transition strength $\alpha$: Defined as the ratio of the latent heat released during the phase transition to the radiation energy density at that time, quantifying the intensity of the phase transition.
	\item Normalized inverse duration of the phase transition $\beta/H_p$: Quantifies the rapidity of the phase transition relative to the Hubble expansion rate at percolation temperature, where $H_p$ denotes the Hubble parameter at the percolation temperature.
	\item Bubble wall velocity $v_w$: 
    The speed at which the boundary of a nucleated bubble of true vacuum expands during a SFOPT. This velocity is a key parameter in early-universe physics, especially in scenarios involving electroweak baryogenesis and GW production. For some EWPT models, this velocity is determined by the balance between the driving force and the plasma friction.
\end{itemize}

The mapping from the fundamental model parameter $\Lambda$ to the set of phase transition parameters constitutes a  nonlinear and computationally intensive process. This mapping represents the first step in the theoretical prediction.

\subsection{Physical Motivation and Limitations of the Model}
The advantage of this model is its connection to UV-complete theories: the dimension-six operator can be derived by integrating out heavy particles from various UV-complete models, such as singlet extensions, composite Higgs models, and two-Higgs-doublet models~\cite{Cao:2017oez}, giving this framework broad theoretical applicability. The phase transition dynamics are governed by a single parameter $\Lambda$, which simplifies the parameter reconstruction procedure. Observations of the GW spectrum can be directly translated into information of $\Lambda$, effectively avoiding complex degeneracies in multi-parameter spaces. Moreover, within reasonable parameter ranges, this model remains consistent with existing electroweak precision measurements.

However, the model faces inherent limitations. The parameter $\Lambda$ is constrained by both Higgs self-coupling measurements and vacuum stability requirements, which limit the extent to which the dimension-six term can modify the barrier structure of 
Higgs potential. As $\Lambda$ increases, the phase transition strength weakens, potentially rendering the GW signal amplitude below detector sensitivity thresholds. The literature estimates a detectable parameter window: $550~\mathrm{GeV} \lesssim \Lambda \lesssim 890~\mathrm{GeV}$~\cite{Huang:2015izx}.
Despite certain limitations, the dimension-six model serves as an ideal representative framework for exploring the EWPT. It incorporates rich possibilities for new physics while maintaining sufficient simplicity to enable quantitative analyses of GW detectability and parameter reconstruction accuracy.

\subsection{Phase transition gravitational wave from sound wave}

GWs from SFOPT are generated through three primary mechanisms: bubble collisions~\cite{Caprini:2015zlo,Huber:2008hg}, SWs~\cite{Hindmarsh:2016lnk,Konstandin:2017sat,Hindmarsh:2019phv,Jinno:2020eqg,Jinno:2022mie}, and turbulence~\cite{Caprini:2009yp,RoperPol:2019wvy}. We focus on the SW mechanism here, as it dominates when EWPT bubble walls reach terminal velocity~\cite{Bodeker:2009qy}. Bubble collisions and turbulence contribute subdominantly at the electroweak scale. The GW production mechanism operates as follows: expanding bubbles collide with each other, creating coherent spherical sound shells in the plasma. These sound shells persist and continue to propagate long after the phase transition completes, colliding and overlapping with shells from other bubbles, rendering SWs an efficient GW source.

Recent hydrodynamic simulations~\cite{Jinno:2022mie,Caprini:2024gyk} show that the GW energy spectrum from SWs can be accurately fitted by a double broken power-law (DBPL) template
\begin{equation}
	\Omega_{\mathrm{GW}}^{\mathrm{SW}}(f) = \Omega_{2}  \left(\frac{f}{f_2}\right)^{n_2} \left[1+\left(\frac{f}{f_1}\right)^{a_1}\right]^{\frac{-n_1+n_2}{a_1}} \left[1+\left(\frac{f}{f_2}\right)^{a_2}\right]^{\frac{-n_2+n_3}{a_2}}.
\end{equation}
The spectral indices of the SW are fixed by physical processes: 
$n_1=3$, $n_2=1$, $n_3=-3$, $a_1=2$, and $a_2=4$~\cite{Jinno:2022mie}. Therefore, only three geometric parameters need to be reconstructed: the amplitude $\Omega_2$, which represents the spectral intensity at the second break frequency, and the frequency break points $f_1$ and $f_2$, which reflect the bubble size and sound shell thickness, respectively.

These geometric parameters to be reconstructed are related to the phase transition thermodynamic parameters through~\cite{Jinno:2022mie,Caprini:2024gyk}
\begin{equation}
\begin{aligned}
f_1 &\simeq 0.2 H_{*,0} (H_p R_p)^{-1},\\ 
f_2 &\simeq 0.5 H_{*,0} \Delta_w^{-1} (H_p R_p)^{-1},\\ 
\Omega_2 &\approx 0.55 \, h^2 F_{\mathrm{GW},0} A_{\mathrm{sw}} K^2 (H_p \tau_{\mathrm{sw}}) (H_p R_p).
\end{aligned}
\end{equation}
Here, $\Delta_w=\xi_{\text{shell}}/\max(v_w, c_s)$ is related to the sound shell thickness, with $\xi_{\text{shell}}=|v_w-c_s|$ representing the velocity difference across the bubble wall, where $c_s$ is the sound speed in the plasma. We adopt $h = 0.67$ from Planck observation~\cite{Planck:2018vyg}. Following Ref.~\cite{Caprini:2019egz}, the scale of bubble collisions is characterized by 
$H_p R_p=(8 \pi)^{1 / 3} \max \left(v_w, c_s\right) H_p/ \beta$, where $R_p$ is the mean bubble separation at percolation temperature. $A_{\mathrm{sw}} \simeq 0.11$ is a fitting constant~\cite{Jinno:2022mie,Caprini:2024gyk}. The kinetic energy fraction $K \simeq 0.6 \kappa \alpha /(1+\alpha)$, where the numerical factor 0.6 arises from efficiency corrections in multi-bubble collision scenarios~\cite{Jinno:2022mie, Caprini:2024gyk}, and $\kappa$ is the kinetic energy fraction for an individual bubble. The $\tau_{\mathrm{sw}}$ is the duration of the SW source
\begin{equation}
H_p \tau_{\mathrm{sw}}=\min \left[\frac{H_p R_p}{\sqrt{\bar{v}_f^2}}, 1\right],
\end{equation} 
where $\bar{v}_f=\sqrt{3K/4}$ is the average fluid velocity~\cite{Caprini:2019egz}. $H_{*, 0}$ denotes the redshifted Hubble parameter
\begin{equation}
H_{*, 0}=1.65 \times 10^{-5}~\mathrm{Hz}\left(g_* / 100\right)^{1 / 6}\left(T_p / 100~\mathrm{GeV}\right),
\end{equation}
where $g_*$ is the effective number of degrees of freedom at the phase transition temperature. 
The observed GW spectrum today  accounts for redshift effects   through the factor $F_{\mathrm{GW}, 0}$, defined as
\begin{equation}
h^2 F_{\mathrm{GW}, 0} \approx 1.64 \times 10^{-5}\left(\frac{100}{g_*}\right)^{1 / 3}.
\end{equation}
Through these relations, the phase transition parameters $\left(T_p, \alpha, \beta / H_p, v_w\right)$ are mapped to the observable spectral parameters $\left(\Omega_2, f_1, f_2\right)$. Bayesian inference reconstructs the former from the latter, thereby constraining the new physics scale $\Lambda$.

\subsection{Theoretical uncertainties of the effective potential}

Before turning to the reconstruction, it is useful to outline the principal sources of uncertainty entering our pipeline, so that the precision quoted in subsequent sections can be interpreted in the proper context. In the present framework, four principal sources contribute: first, theoretical uncertainties associated with the perturbative treatment of the finite-temperature effective potential; second, the uncertainty in the bubble wall velocity, to which the GW spectrum is highly sensitive; third, the uncertainty associated with the machine-learning inference procedure; and fourth, the statistical uncertainty arising from detector noise and astrophysical foregrounds. In this section, we focus on the simple discussions of the first source, while the remaining three are discussed in subsequent sections.

In this work, we adopt the on-shell renormalization scheme, in which the renormalization conditions are imposed at the physical Higgs mass and vacuum expectation value, thereby fixing the renormalization point near the electroweak scale. Residual theoretical uncertainty nevertheless remains due to the perturbative truncation of the effective potential and is conventionally estimated by varying the renormalization scale $\bar\mu$ in an $\overline{\rm MS}$-type treatment. Variations of $\bar\mu$ over the range $(0.5\text{--}2\pi)T$ have been reported to induce $20$--$30\%$ shifts in $T_c$, $200$--$800\%$ shifts in $\alpha$, and $40$--$200\%$ shifts in $\beta/H$, corresponding to $\frac{\Delta\Omega_{\rm GW}}{\Omega_{\rm GW}} = \mathcal{O}(10^2\text{--}10^3)$,
thereby providing an indicative estimate of the size of missing higher-order corrections~\cite{Croon:2020cgk}.

A second important source of uncertainty arises from the gauge dependence of the one-loop effective potential. In conventional four-dimensional (4d) perturbative calculations, the effective potential evaluated is generally gauge dependent, leading to ambiguities in the predicted phase-transition parameters and GW observables~\cite{Patel:2011th,Garny:2012cg}. Some approaches, such as strict $\hbar$-expansion methods or dimensionally reduced 3d effective theories, can substantially reduce this uncertainty~\cite{Croon:2020cgk,Lewicki:2024xan}, but lie beyond the scope of the present work.

A further source of uncertainty comes from different resummation schemes arising from infrared divergences associated with bosonic Matsubara zero modes. We adopt the Parwani scheme, in which the field-dependent masses are replaced by thermal masses throughout the one-loop integrals. An alternative approach is the Arnold--Espinosa scheme, in which the daisy contribution is added as a separate term. These two schemes differ formally at higher orders in the perturbative expansion (typically beyond $\mathcal{O}(g^4)$) and might yield numerically distinct predictions for the phase-transition parameters~\cite{Croon:2020cgk,Cho:2021itv}. Other resummation schemes could also make different predictions~\cite{Bittar:2025lcr}.

Beyond the one-loop uncertainties discussed above, the truncation of the perturbative expansion at one-loop order introduces further higher-order corrections. Taken together, the renormalization-scale dependence, gauge dependence, resummation-scheme dependence, and perturbative truncation uncertainty can all be substantial within the conventional 4d perturbative framework. While alternative approaches such as the dimensionally reduced 3d effective theory can mitigate some of these theoretical systematics~\cite{Lewicki:2024xan}, a systematic reduction of these uncertainties is required before the statistical precision quoted in our reconstruction can be translated into a robust physical constraint on the underlying new-physics scale. Addressing these theoretical uncertainties lies beyond the scope of the present work and is left for future study.

\section{Mapping Model Parameters to Geometric Parameters and the Physical Origin of Degeneracy}\label{sec:map}

The previous section described the forward mapping from the dimension-six model parameters to the geometric parameters
\begin{equation}
\Lambda  \xrightarrow{} V_{\text{eff}}(h_0, T) \xrightarrow{\text {}}\left(T_p, \alpha, \beta / H_p, v_w\right)\xrightarrow{\text {}} \left(\Omega_2, f_1, f_2\right)
\end{equation}
This represents a critical step in linking phase transition physics to GW observable signals—namely, the formulation of the forward problem. 
The analysis is structured in three sequential stages: detection of the signal, quantitative characterization of its features, and reconstruction of model parameters to uncover the underlying physical mechanisms. This reconstruction involves parameters at multiple levels: geometric parameters that describe the spectral shape ($\Omega_2$, $f_1$, $f_2$), phase transition parameters that characterize the thermodynamics $\left(T_p, \alpha, \beta / H_p, v_w\right)$, and ultimately the model parameter in the Lagrangian $(\Lambda)$.
The geometric parameters are determined from the phase transition parameters through the following analytical relations:

\begin{itemize} 
\item \textbf{Break frequencies}: The two frequency breaks encode distinct physical scales of the transition. The lower break $f_1$ reflects the mean bubble spacing $R_p$, setting the macroscopic characteristic scale; faster transitions (larger $\beta / H_p$) shift $f_1$ to higher frequencies. The upper break $f_2$ traces the sound shell width $\Delta_w$, governed mostly by the wall velocity $v_w$, thus capturing the microphysics of bubble expansion.
\item \textbf{Amplitude}: The GW energy scales with both the available kinetic energy $K$ and the active source time $\tau_{\mathrm{sw}}$. The latter is set by whichever terminates first: Hubble expansion ($1/H_p$) or turbulent decay ($ R_p/\sqrt{\bar{v}_f^2}$), which produces $\tau_{\mathrm{sw}} = \min[ R_p/\sqrt{\bar{v}_f^2}, 1/H_p]$. This minimum operation creates a nonlinearity in the $(\alpha, \beta/H_p) \to \Omega_2$ mapping, as two competing physical processes control the source lifetime. Such nonlinearity introduces degeneracies in Bayesian parameter reconstruction: distinct thermodynamic configurations $(\alpha, \beta/H_p)$ can produce identical amplitudes $\Omega_2$.
\end{itemize}

This mapping relation reveals the physical origin of parameter degeneracies. We attempt to determine four independent thermodynamic parameters $(T_p, \alpha, \beta / H_p, v_w)$ from three independent geometric observables (which can be expressed as $(\Omega_2, f_1, f_2)$). Information is lost in the projection from the physical space to the observable space.

We consider two distinct strategies for reconstructing the model parameters from GW observables:

\textbf{Strategy 1: Two-step mapping.} The first approach faithfully follows the complete physical chain:
$$\left(\Omega_2, f_1, f_2\right) \xrightarrow{\text {  }}\left(T_p, \alpha, \beta / H_p, v_w\right) \xrightarrow{\text { }} \Lambda$$
The advantage of this approach is its clear physical interpretation: it can simultaneously constrain all phase transition thermodynamic parameters. The disadvantage is that each mapping step accumulates errors and uncertainties, resulting in a high computational cost. Moreover, the degeneracies among thermodynamic parameters can lead to degenerate posterior distributions.

\textbf{Strategy 2: Direct mapping.} The second approach establishes a direct reconstruction from $\Lambda$ to the geometric parameters:
$$
\left(\Omega_2, f_1, f_2\right) \xrightarrow{\text {   }} \Lambda
$$
The implementation proceeds as follows: (i) grid-sample $\Lambda$ values across parameter space; (ii) run CosmoTransitions~\cite{Wainwright:2011kj} and the GW spectrum calculation for each $\Lambda$ to establish the mapping $\Lambda \to (\Omega_2, f_1, f_2)$; (iii) construct an interpolation function; and (iv) perform Bayesian inference directly on $\Lambda$. 

The advantages of this approach are: (1) it avoids multi-step error propagation; (2) it eliminates the need to solve the inverse problem at each likelihood evaluation; and (3) it produces a simpler posterior distribution---the posterior on $\Lambda$ is much less complex than the joint posterior on $(T_p, \alpha, \beta / H_p, v_w)$. The limitation is that this approach cannot directly constrain the thermodynamic parameters and requires extensive precomputation of forward models across the parameter space.

Based on these considerations, this study adopts Strategy 2. We establish a direct mapping from the observed GW geometric parameters to $\Lambda$ by training an interpolation function on a grid of precomputed forward models. The training set generation procedure is as follows:
\begin{enumerate}
	\item \textbf{Grid sampling}: Uniformly sample the $\Lambda$ parameter space ($548~\mathrm{GeV} \lesssim \Lambda \lesssim 560~\mathrm{GeV}$).
	\item \textbf{Forward computation}: For each sampled $\Lambda$ value, run CosmoTransitions to compute the phase transition parameters, then calculate $(\Omega_2, f_1, f_2)$ using the DBPL template.
	\item \textbf{Feature reduction}: With the wall velocity fixed, the ratio $f_2/f_1 $ becomes approximately constant. Since $f_1$ and $f_2$ are linearly related, they provide only one independent constraint rather than two. We therefore reduce the feature space to $(\Omega_2, f_2)$, retaining $f_2$ as it is more sensitive to the frequency band of detectors.
\end{enumerate}

Although the parameter space is formally one-dimensional in \(\Lambda\), the inverse mapping from \((\Omega_{2}, f_{2})\) to \(\Lambda\) is highly nonlinear because the phase-transition dynamics depend on \(\Lambda\) in a complicated way. Simple inversion strategies—such as bisection or linear interpolation—tend to be unstable and lead to large reconstruction errors. Instead of relying on such direct interpolation, we employ an ensemble machine-learning framework~\cite{Breiman:2001hzm,hastie2009elements,sivia2006data}, in which multiple regression models (including Gaussian-process regression, random forest, gradient-boosted trees, and a multi-layer perceptron) are trained on simulated data and combined to yield robust predictions for previously unseen inputs.

\section{Parameter reconstruction pipeline of TianQin}\label{sec:noise}
\subsection{Noise model of TianQin}

To extract faint GW signals from dominant noise sources, the TianQin mission employs the technique of Time Delay Interferometry (TDI)~\cite{PhysRevD.63.021101,Hogan:2001jn,PhysRevD.65.082003}. TDI constructs a set of virtual interferometric channels by applying specific time-delayed combinations to the raw laser link data. This approach effectively suppresses laser noise, which would otherwise overwhelm the GW signal. The construction of these channels follows a hierarchical structure. It begins with the fundamental one-way and round-trip laser links between spacecraft, which are then combined through differential operations to form Michelson channels. These intermediate channels serve as building blocks for the final TDI observables, which are optimized for noise cancellation and signal extraction in the space-based interferometric configuration.

In the preliminary design of the TianQin mission, three TDI channels---denoted as X, Y, and Z---are constructed using the spacecraft nodes $A_0$, $B_0$, and $C_0$ as the centers of interference. These channels are combined to form the AET basis~\cite{Vallisneri:2012np,tinto2021time,PhysRevD.106.124027}
\begin{equation}
\mathrm{A}=\frac{1}{\sqrt{2}}(\mathrm{Z-X}),\quad \mathrm{E}=\frac{1}{\sqrt{6}}(\mathrm{X-2 Y+Z}),\quad \mathrm{T}=\frac{1}{\sqrt{3}}(\mathrm{X+Y+Z}).
\end{equation}

\begin{figure}[h!]
	\centering
	\includegraphics[width=0.7\linewidth]{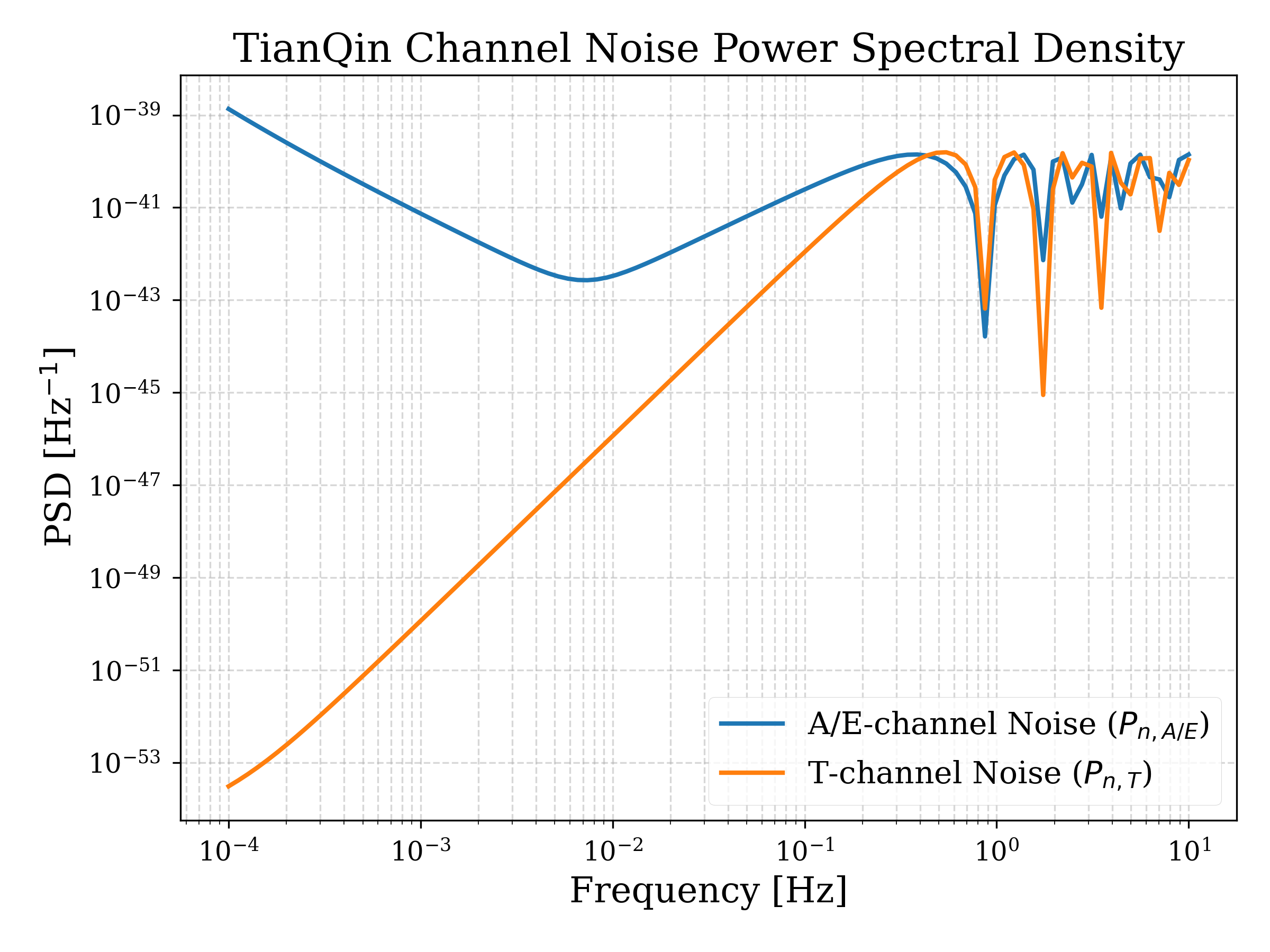}
	\caption{Noise PSD of the AET channels in the TianQin detector. These orthogonal TDI channels are constructed to suppress laser frequency noise across the mission's frequency band.}
	\label{fig:tqnoi}
\end{figure}

In the AET channel configuration, the A and E channels share an identical expression for their noise PSD~\cite{Liang:2022ufy}
\begin{equation}
P_{nA}(f)=\frac{2 \sin ^2\left[f / f_*\right]}{L^2}\left[\left(\cos \left[f / f_*\right]+2\right) S_p(f)+2\left(\cos \left[2 f / f_*\right]+2 \cos \left[f / f_*\right]+3\right) \frac{S_a(f)}{(2 \pi f)^4}\right].
\end{equation}

Here, $S_p(f)$ denotes the position noise, and $S_a(f)$ represents the acceleration noise (see Tab.~\ref{tab:pri}). The characteristic frequency is defined as $f_* = c / (2\pi L)$, where $L$ is the interferometer arm length and $c$ is the speed of light. The noise PSD for the T channel is given by
\begin{equation}
P_{n T}(f)=\frac{8 \sin ^2\left[f / f_*\right] \sin ^2\left[f /\left(2 f_*\right)\right]}{L^2}\left[S_p(f)+4 \sin ^2\left[f /\left(2 f_*\right)\right] \frac{S_a(f)}{(2 \pi f)^4}\right].
\end{equation}

Figure~\ref{fig:tqnoi} shows the instrumental noise PSD for the AET channels in the TianQin detector. The blue curve represents the noise PSD of the A and E channels, $P_{nA/E}$, which exhibit identical noise characteristics. The orange curve corresponds to the T channel noise spectrum, $P_{nT}$. Unlike the A/E channels, the T channel shows significant suppression of noise in the low-frequency regime. Within TianQin's core sensitivity band, the noise level of the T channel is substantially lower than that of the A/E channels.

\subsection{Response Functions of AET channels}

\begin{figure}[h!]
    \centering
    \includegraphics[width=0.5\linewidth]{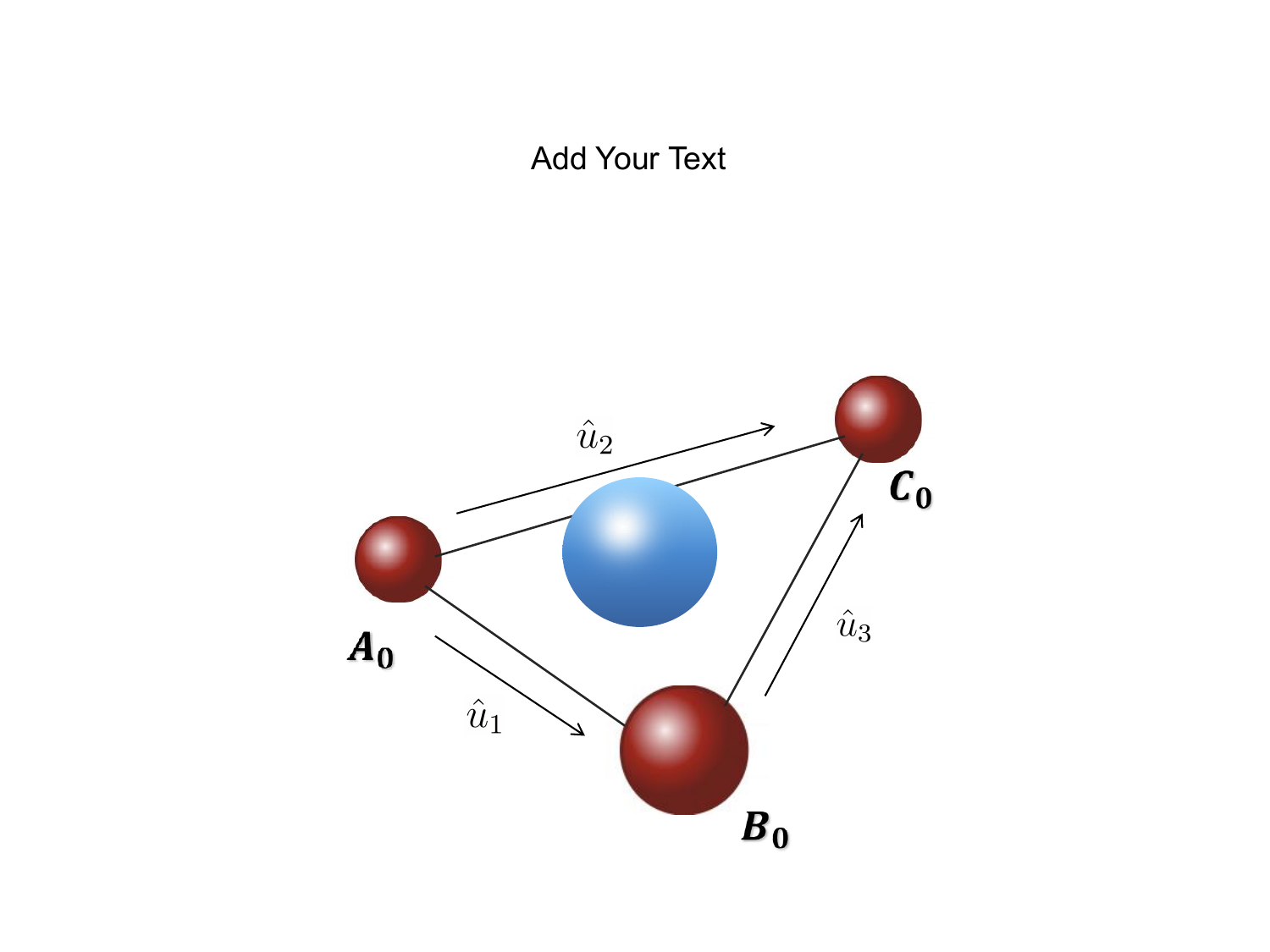}
    \caption{Equal-arm Michelson channels of the regular triangle detector.}
    \label{fig:place}
\end{figure}

Figure~\ref{fig:place} illustrates the construction of equal-arm Michelson channels using one vertex satellite of the triangular configuration and its two adjacent arms. By selecting each corner satellite—$A_0$, $B_0$, and $C_0$—as the interferometric vertex, a set of Michelson channels $\mathrm{M}_1$, $\mathrm{M}_2$, and $\mathrm{M}_3$ can be formed.

Each Michelson channel is constructed from two distinct round-trip links, where a round-trip link is defined as the complete propagation of a laser signal along a single arm and back. Introducing the arm vectors as $\overrightarrow{A_0 B_0} = L \hat{u}_1$, $\overrightarrow{A_0 C_0} = L \hat{u}_2$, and $\overrightarrow{B_0 C_0} = L \hat{u}_3$, the response function of the Michelson channel set can be expressed as the differential response between the two round-trip links~\cite{Liang:2022ufy}
\begin{equation}
\begin{aligned}
F_{\mathrm{M}_1}^P\left(f, \hat{k}, t_0\right) & =F_{\mathrm{II}}^P\left[f, \hat{k}, \hat{u}_1\left(t_0\right)\right]-F_{\mathrm{II}}^P\left[f, \hat{k}, \hat{u}_2\left(t_0\right)\right], \\
F_{\mathrm{M}_2}^P\left(f, \hat{k}, t_0\right) & =F_{\mathrm{II}}^P\left[f, \hat{k}, \hat{u}_3\left(t_0\right)\right]-F_{\mathrm{II}}^P\left[f, \hat{k},-\hat{u}_1\left(t_0\right)\right], \\
F_{\mathrm{M}_3}^P\left(f, \hat{k}, t_0\right) & =F_{\mathrm{II}}^P\left[f, \hat{k},-\hat{u}_2\left(t_0\right)\right]-F_{\mathrm{II}}^P\left[f, \hat{k},-\hat{u}_3\left(t_0\right)\right].
\end{aligned}
\end{equation}
Here, $F_{\mathrm{II}}^P$ denotes the response function of a round-trip link, with the negative sign indicating the reversed direction of the arm. The unit vector $\hat{k}$ represents the propagation direction of the GW from the source to the detector.
These Michelson channels serve as building blocks for constructing TDI channels through time-delayed combinations, which effectively suppress laser frequency noise. The XYZ channel set is formed by differencing the instantaneous response of the Michelson channels with their delayed counterparts, offset by $2L/c$. Taking the X channel as an example, its response function is composed of the $\mathrm{M}_1$ channel evaluated at different time instances
\begin{equation}
F_{\mathrm{X}}^P\left(f, \hat{k}, t_0\right)=F_{\mathrm{M}_1}^P\left(f, \hat{k}, t_0\right)-F_{\mathrm{M}_1}^P\left(f, \hat{k}, t_0-2 L / c\right)=\left(1-e^{-i 2 f / f_*}\right) F_{\mathrm{M}_1}^P\left(f, \hat{k}, t_0\right).
\end{equation}
The delay factor $\left(1 - e^{-i 2 f / f_*}\right)$ arises from the phase accumulation in the frequency domain due to the time delay $2L/c$. Following the cyclic permutation principle, the Y and Z channels are constructed analogously from the $\mathrm{M}_2$ and $\mathrm{M}_3$ Michelson channels, respectively
\begin{equation}
\begin{aligned}
& F_{\mathrm{Y}}^P\left(f, \hat{k}, t_0\right)=\left(1-e^{-i 2 f / f_*}\right) F_{\mathrm{M}_2}^P\left(f, \hat{k}, t_0\right), \\
& F_{\mathrm{Z}}^P\left(f, \hat{k}, t_0\right)=\left(1-e^{-i 2 f / f_*}\right) F_{\mathrm{M}_3}^P\left(f, \hat{k}, t_0\right).
\end{aligned}
\end{equation}
The correlation of AET channel responses to the SGWB is characterized by the overlap reduction function (ORF). For an isotropic background, the ORF is the sky-averaged geometric correlation factor~\cite{Liang:2022ufy}
\begin{equation}
\Gamma_{I J}(f, t)=\frac{1}{4 \pi} \int_{S^2} \mathrm{d}^2 \hat{\Omega}_{\hat{k}} \Upsilon_{I J}(f, \hat{k}, t),
\end{equation}
The term $\mathrm{d}^2 \hat{\Omega}_{\hat{k}}$ denotes the differential solid angle element in the direction $\hat{k}$. The geometric correlation factor $\Upsilon_{IJ}(f, \hat{k}, t)$ quantifies the correlated response of channels $I$ and $J$ to GWs propagating from direction $\hat{k}$
\begin{equation}
\Upsilon_{I J}(f, \hat{k}, t)=\frac{1}{2} \sum_{P=+, \times} F_I^P(f, \hat{k}, t) F_J^{P *}(f, \hat{k}, t) e^{-i 2 \pi f \hat{k} \cdot\left[\vec{x}_I(t)-\vec{x}_J(t)\right] / c},
\end{equation}
where $\vec{x}_I(t)$ and $\vec{x}_J(t)$ denote the lasers interference sites of channel $I$ and $J$. 
The expression involves a summation over the two polarization states, $+$ and $\times$, while the phase term accounts for the phase difference arising from the spatial separation between the two channels. As a special case, the transfer function of a single channel corresponds to the autocorrelation of its own response function
\begin{equation}
\mathcal{R}_I(f)=\frac{1}{8 \pi} \sum_{P=+, \times} \int_{S^2} \mathrm{d}^2 \hat{\Omega}_{\hat{k}} F_I^P(f, \hat{k}, t) F_I^{P *}(f, \hat{k}, t)=\Gamma_{I I}(f).
\end{equation}

Having established the XYZ channel responses, the AET channels are now 
constructed through linear combinations. Taking the A channel as an example, its response function can be expressed as the normalized difference between the responses of the Z and X channels
\begin{equation}
F_A^P(f, \hat{k}, t)= \left[F_Z^P(f, \hat{k}, t) e^{-i 2 \pi f \hat{k} \cdot \overrightarrow{A_0 C_0}(t) / c}-F_X^P(f, \hat{k}, t)\right] / \sqrt{2}.
\end{equation}
From this relation, one can derive that the transfer functions of the AET channels satisfy
\begin{equation}
\begin{aligned}
\mathcal{R}_{\mathrm{A}}(f) & =\frac{1}{8 \pi} \sum_{P=+, \times} \int_{S^2} \mathrm{d}^2 \hat{\Omega}_{\hat{k}} F_{\mathrm{A}}^P(f, \hat{k}, t) F_{\mathrm{A}}^{P *}(f, \hat{k}, t) \\
& =\frac{1}{2}\left[\mathcal{R}_{\mathrm{X}}(f)+\mathcal{R}_{\mathrm{Z}}(f)-2 \Gamma_{\mathrm{XZ}}(f)\right] \\
& =\mathcal{R}_{\mathrm{X}}(f)-\Gamma_{\mathrm{XY}}(f), \\
\mathcal{R}_{\mathrm{E}}(f) & =\frac{1}{8 \pi} \sum_{P=+, \times} \int_{S^2} \mathrm{d}^2 \hat{\Omega}_{\hat{k}} F_{\mathrm{E}}^P(f, \hat{k}, t) F_{\mathrm{E}}^{P *}(f, \hat{k}, t) \\
& =\frac{1}{6}\left[\mathcal{R}_{\mathrm{X}}(f)+4 \mathcal{R}_{\mathrm{Y}}(f)+\mathcal{R}_{\mathrm{Z}}(f)-4 \Gamma_{\mathrm{XY}}(f)+2 \Gamma_{\mathrm{XZ}}(f)-4 \Gamma_{\mathrm{YZ}}(f)\right] \\
& =\mathcal{R}_{\mathrm{X}}(f)-\Gamma_{\mathrm{XY}}(f), \\
\mathcal{R}_{\mathrm{T}}(f) & =\frac{1}{8 \pi} \sum_{P=+, \times} \int_{S^2} \mathrm{d}^2 \hat{\Omega}_{\hat{k}} F_{\mathrm{T}}^P(f, \hat{k}, t) F_{\mathrm{T}}^{P *}(f, \hat{k}, t) \\
& =\frac{1}{3}\left[\mathcal{R}_{\mathrm{X}}(f)+\mathcal{R}_{\mathrm{Y}}(f)+\mathcal{R}_{\mathrm{Z}}(f)+2 \Gamma_{\mathrm{XY}}(f)+2 \Gamma_{\mathrm{XZ}}(f)+2 \Gamma_{\mathrm{YZ}}(f)\right]
\\& =\mathcal{R}_{\mathrm{X}}(f)+2\Gamma_{\mathrm{XY}}(f).
\end{aligned}
\end{equation}
The AET transformation achieves complete orthogonalization of the channel responses. Calculations confirm that
\begin{equation}
\Gamma_{AE}(f) = \Gamma_{AT}(f) = \Gamma_{ET}(f) = 0,
\end{equation}
indicating that the three channels respond to the SGWB in a statistically independent manner, effectively eliminating inter-channel signal correlations. The following relations are thus obtained
\begin{equation}
\begin{aligned}
&\begin{aligned}
& \mathcal{R}_{\mathrm{A}}(f)=\mathcal{R}_{\mathrm{E}}(f)=\frac{3}{2} \mathcal{R}_{\mathrm{X}}(f), \\
& \mathcal{R}_{\mathrm{T}}(f)=o\left(\mathcal{R}_{\mathrm{X}}(f)\right), \quad f \ll f_*.
\end{aligned}
\end{aligned}
\end{equation}
\begin{figure}[h!]
	\centering
	\includegraphics[width=0.7\linewidth]{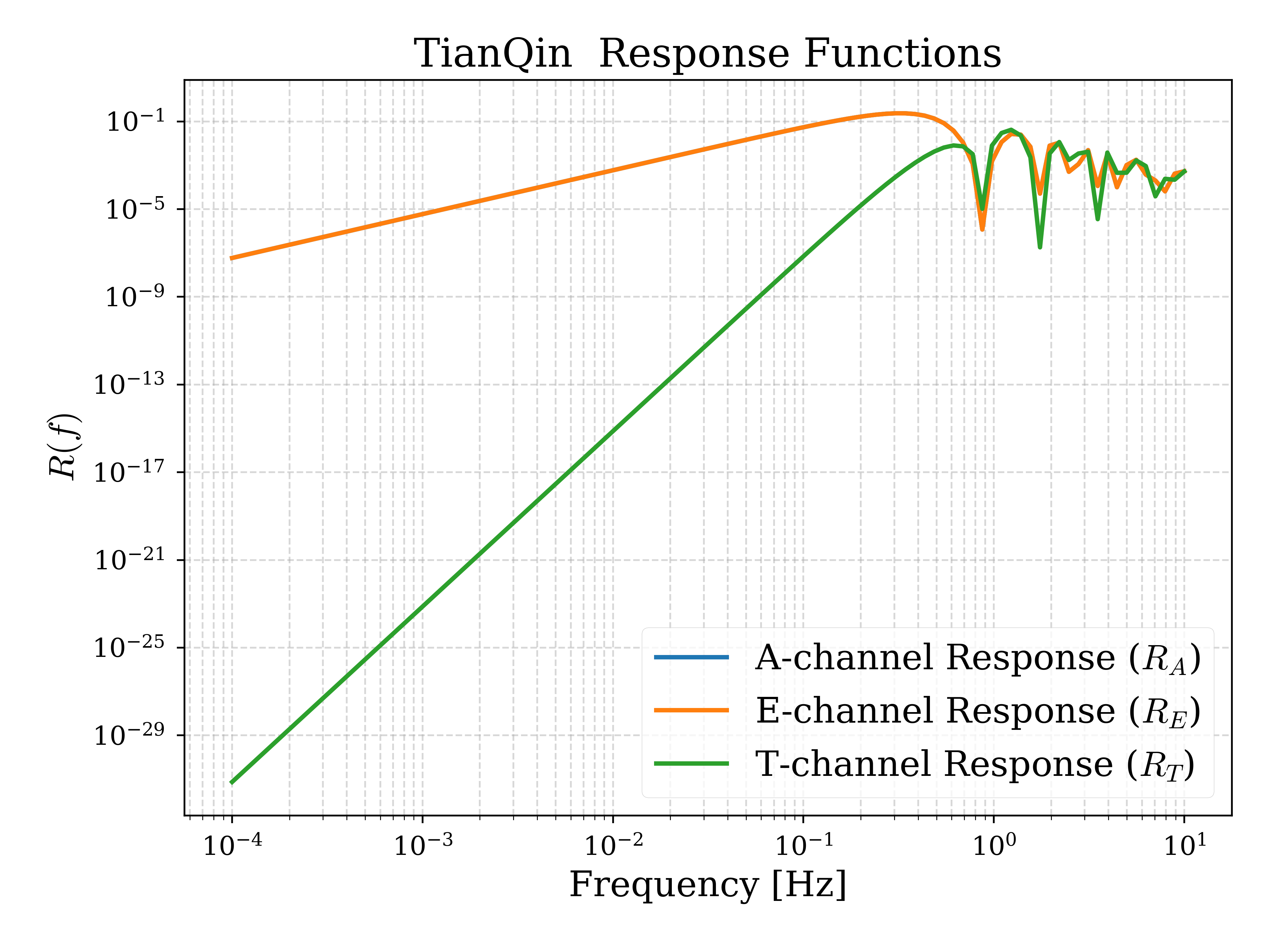}
	\caption{Response Functions of AET channels in TianQin.}
	\label{fig:tqres}
\end{figure}

Figure~\ref{fig:tqres} shows the response functions for TianQin's AET channels. The A (blue) and E (orange) channels exhibit strong, uniform sensitivity across the frequency band and serve as primary science channels. The T channel (green) has lower sensitivity, particularly at low frequencies, making it suitable for noise monitoring and calibration rather than signal detection. 
\begin{figure}[h!]
	\centering
	\includegraphics[width=0.7\linewidth]{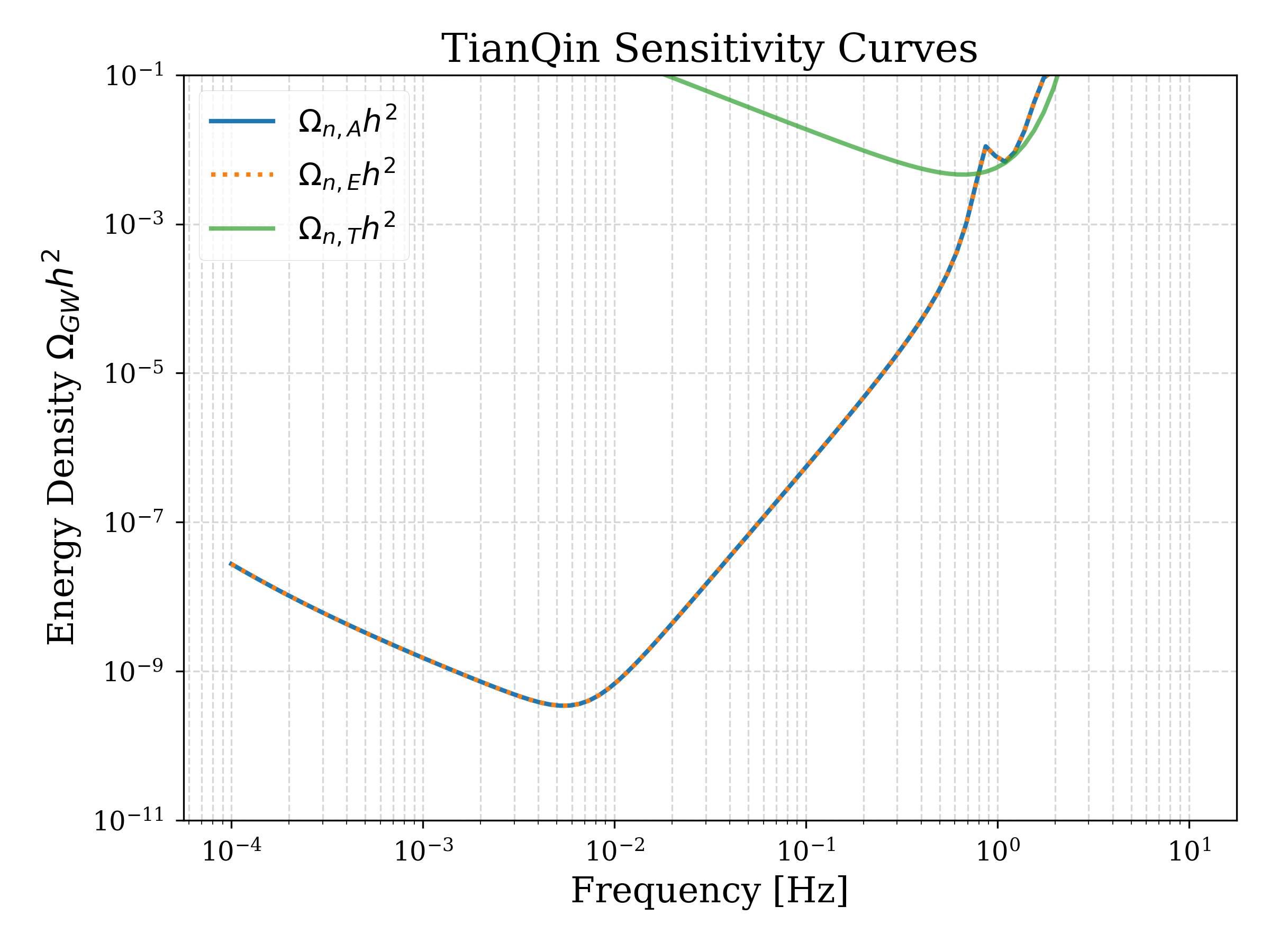}
	\caption{Sensitivity curves of the AET channels for the TianQin detector.}
	\label{fig:tqsen}
\end{figure}

Figure~\ref{fig:tqsen} illustrates the TianQin sensitivity in the AET channel basis, expressed in terms of the GW energy density parameter $\Omega_{\mathrm{GW}} h^2$. The sensitivity curves for the A channel ($\Omega_{n, A} h^2$, solid blue line) and the E channel ($\Omega_{n, E} h^2$, dashed orange line) are indistinguishable, indicating that both channels exhibit comparable sensitivity to the GW energy density. In contrast, the T channel ($\Omega_{n, T} h^2$, solid green line) shows significantly reduced sensitivity at low frequencies and features pronounced oscillations at high frequencies. The energy density sensitivity is computed as the ratio of the strain noise PSD to the corresponding response function
\begin{equation}\label{eq:ed}
\Omega_{n, i } h^2(f)=\frac{4 \pi^2 f^3}{3\left(H_0 / h\right)^2}S_{n, i }(f)=\frac{4 \pi^2 f^3}{3\left(H_0 / h\right)^2} \frac{P_{n,i }(f)}{\mathcal{R}_{i }(f)},
\end{equation}
where $i \in \{A, E, T\}$ denotes the TianQin channel, and $H_0 / h \simeq 3.24 \times 10^{-18} \mathrm{~s}^{-1}$ is the normalized Hubble parameter.

\subsection{Two types of foregrounds}
In addition to instrumental noise and the possible phase transition GW signals, TianQin's observational data contain a stochastic foreground arising from unresolved astrophysical sources. The dominant contributions to this foreground originate from two classes of compact binary systems: extragalactic binaries and Galactic compact binaries.

The extragalactic foreground is composed of the superposition of GW emitted during the inspiral phase of numerous unresolved stellar-mass black hole binaries and neutron star binaries distributed across cosmological scales. This cumulative signal manifests in the TianQin frequency band as an approximately isotropic stochastic background. Theoretical models predict that its spectral shape follows a power-law behavior~\cite{Phinney:2001di,Regimbau:2011rp,Perigois:2020ymr,Babak:2023lro,Lehoucq:2023zlt}
\begin{equation}
\Omega_{\mathrm{GW}}^{\mathrm{Ext}}h^2(f) = h^2 \Omega_{\mathrm{Ext}} \left(\frac{f}{f_{\mathrm{ref}}}\right)^{2/3},
\end{equation}
where the exponent of $2 / 3$ arises from the Post-Newtonian approximation of the binary systems during their inspiral phase. The normalization frequency is set at $f_{\mathrm{ref}}=10^{-3} \mathrm{~Hz}$. The value of the amplitude $\Omega_{\text{Ext}}$ is highly dependent on factors such as the cosmological redshift evolution, the initial mass distribution, and thus is currently subject to significant theoretical uncertainty.

We adopt an estimated fiducial value of $\log_{10}\left(h^2 \Omega_{\mathrm{Ext}}\right) = -12.38$~\cite{Babak:2023lro}. To robustly incorporate current observational constraints and their associated uncertainties into the parameter estimation process, we employ a Gaussian prior $\mathcal{N}\left(-12.38,\, 0.17^2\right)$, effectively anchoring the inference near existing bounds on the cosmological GW background. Here, $\mathcal{N}(\mu, \sigma^2)$ denotes a Gaussian distribution with mean $\mu$ and variance $\sigma^2$.

In the field of space-based GW detection---particularly in the milli-hertz frequency band---the population of compact binaries within the Milky Way, especially double white dwarfs (DWDs), constitutes an unavoidable and highly intense astrophysical foreground. Accurate modeling and separation of this foreground are essential for extracting cosmological signals and resolving individual Galactic sources.

This study focuses on the time-averaged signal obtained by integrating over the entire sky throughout the full observation period $T_{\mathrm{obs}}$. The Galactic foreground energy density spectrum model adopted here follows the formulation presented by Karnesis et al.~\cite{Karnesis:2021tsh}, and is expressed as
\begin{equation}
\Omega_{\mathrm{GW}}^{\mathrm{Gal}}h^2(f) = h^2 \Omega_{\mathrm{Gal}}  \frac{f^3}{2} \left(\frac{f}{1~\mathrm{Hz}}\right)^{-7/3} \left[1 + \tanh\left(\frac{f_{\mathrm{knee}} - f}{f_d}\right)\right] e^{-(f/f_c)^{\nu}}.
\end{equation}
Here, $\nu = 1.56$ and $f_d = 6.7 \times 10^{-4}~\mathrm{Hz}$. The characteristics of the Galactic foreground are closely tied to the duration of the observation, $T_{\mathrm{obs}}$, as the observation time determines which binary systems can be individually resolved. In this model, two key frequency parameters depend explicitly on $T_{\mathrm{obs}}$

\begin{enumerate}
\item \textbf{Knee frequency}   
($f_{\text{knee}}$): This parameter marks the frequency turnover point from the confusion-dominated regime to the resolved-source-dominated regime.
\begin{equation}
\log _{10}\left(f_{\text {knee}} / \mathrm{Hz}\right)=-0.37 \log _{10}\left(T_{\text {obs}} / \text {year}\right)-2.49.
\end{equation}

\item \textbf{Cutoff frequency} ($f_c$): This parameter controls the location of the high-frequency astrophysical cutoff.
\begin{equation}
\log _{10}\left(f_c / \mathrm{Hz}\right)=-0.15 \log _{10}\left(T_{\text{obs}} / \text {year}\right)-2.72.
\end{equation}
\end{enumerate}
 
The value for the amplitude is set at $\log _{10}\left(h^2 \Omega_{\mathrm{Gal}}\right)=-7.84$. A Gaussian prior $\mathcal{N}\left(-7.84,0.21^2\right)$  is adopted in the parameter estimation~\cite{Karnesis:2021tsh}.

In the Bayesian analysis, the amplitudes of the two foreground components, $\left(\Omega_{\mathrm{Ext}}, \Omega_{\mathrm{Gal}}\right)$, are treated as additional free parameters and jointly inferred alongside the parameters of the cosmological SGWB. This approach offers the advantage of naturally propagating foreground uncertainties into the posterior distributions of cosmological parameters through joint sampling of the parameter space. Constraints from ground-based detectors and Galactic population models are incorporated via prior distributions.
\begin{table}[h]
\centering
\caption{Prior distributions for all reconstructed parameters.}
\begin{tabular}{cc}
    \hline\hline
    parameter & prior \\
    \hline\hline
    $\log_{10}(h^2 \Omega_{\mathrm{Ext}})$ & $\mathcal{N}(-12.38, 0.17^2)$ \\
    $\log_{10}(h^2 \Omega_{\mathrm{Gal}})$ & $\mathcal{N}(-7.84, 0.21^2)$ \\
    $\log_{10}(h^2 \Omega_2)$ & $\mathcal{U}(-14.0, -8.0)$ \\
    $\log_{10}(f_2 / \mathrm{Hz})$ & $\mathcal{U}(-5.0, -1.0)$ \\
    $\log_{10}(f_2 / f_1)$ & $\mathcal{U}(0.5, 3.0)$ \\
    \hline\hline
\end{tabular}\label{tab:pri}
\end{table}

Table~\ref{tab:pri} summarizes the prior distributions for all reconstructed parameters.   $\mathcal{U}(a, b)$ denotes a uniform distribution over the interval $[a, b]$. The Galactic and extragalactic foreground amplitudes use Gaussian priors, while the geometric parameters use uniform priors.

\subsection{Data Generation Pipeline}

The simulated data are generated using the PSD of the SGWB and instrumental noise, following the procedure described below. Template-based signal analysis enables us to leverage known theoretical models for more efficient signal extraction.

For our simulation, we assume a total effective observation time of $T_{\text{obs}}=4$ years, divided into $N_{\mathrm{c}}=127$ data segments, each with a duration of $\tau=11.5$ days.  For each data segment $l=1, \ldots, N_{\mathrm{c}}$, the code generates data with a frequency resolution of $\Delta f=1 / \tau \simeq 1.0 \times 10^{-6} \mathrm{~Hz}$. The analysis frequency band covers TianQin's sensitive range of $\left[10^{-4}, 1\right]~\mathrm{Hz}$, yielding approximately $N_f=\left(f_{\max }-f_{\min }\right) / \Delta f \simeq 10^6$ data points per segment. Subsequently, logarithmic binning is applied to compress the data for Bayesian analysis.\\

\textbf{Step 1: Simulated Data Generation in the Frequency Domain}\\

The simulated data are generated directly in the frequency domain. Assuming that the signal and noise are statistically uncorrelated, the data are constructed as
\begin{equation}\label{eq:di}
d_i(f)=\sum_\zeta 
n_i^\zeta(f)+\sum_\eta 
s_i^\eta(f),
\end{equation}
where the summations run over different noise sources (indexed by $\zeta$) and signal components (indexed by $\eta$), respectively. For this study, we consider two types of instrumental noise—arising from TM and OMS—and three signal components: extragalactic and Galactic astrophysical foregrounds, and the primordial SGWB from EWPT. For each data segment \(l\) and TDI channels \(i, j\), we have~\cite{Caprini:2019pxz}
\begin{equation} \label{eq:Di}
 D_{ij}^{(l)}(f) = d_{i}^{(l)}(f) \, d_{j}^{(l)*}(f).
\end{equation}

Under the Gaussian assumption for both signal and noise, we construct simulated data following Ref.~\cite{Flauger:2020qyi}
\begin{equation}
\begin{aligned}
	& S_i(f_i)=\left|\frac{G_{i 1}\left(0, \sqrt{ \Omega_{\mathrm{tot}}h^2\left(f_i\right)}\right)+i G_{i 2}\left(0, \sqrt{ \Omega_{\mathrm{tot}}h^2\left(f_i\right)}\right)}{\sqrt{2}}\right|^2, \\
	& N_i(f_i)=\left|\frac{G_{i 3}\left(0, \sqrt{ \Omega_{\mathrm{n}}h^2\left(f_i\right)}\right)+i G_{i 4}\left(0, \sqrt{ \Omega_{\mathrm{n}}h^2\left(f_i\right)}\right)}{\sqrt{2}}\right|^2.
\end{aligned}
\end{equation}

In this expression, $G_{ij}(\mu, \sigma)$ ($j=1,\ldots,4$) are independent random draws from a Gaussian distribution with mean $\mu$ and standard deviation $\sigma$. These construct the real and imaginary parts of the complex Fourier coefficients for signal and noise. The total signal energy density is $\Omega_{\mathrm{tot}}h^2\left(f_i\right)=\Omega_{\mathrm{GW}}^{\mathrm{SW}}h^2\left(f_i\right)+\Omega_{\mathrm{GW}}^{\mathrm{Ext}}h^2\left(f_i\right)+\Omega_{\mathrm{GW}}^{\mathrm{Gal}}h^2\left(f_i\right)$.  From these, we obtain $d_{i}^{(l)}(f)$, which is then used to compute $D_{ij}^{(l)}(f)$ according to Eqs.~\eqref{eq:di} and~\eqref{eq:Di}.\\

\textbf{Step 2: Binning of Simulated Data in the Frequency Domain}\\
To improve computational efficiency, we reduce the frequency resolution through logarithmic binning. First, we average all data segments
\begin{equation}
\bar{D}_{ij}(f) \equiv \frac{1}{N_{\mathrm{c}}} \sum_{l=1}^{N_{\mathrm{c}}} D_{ij}^{(l)}(f).
\end{equation}
The variance is given by
\begin{equation}
\sigma_{ij}^2(f) = \frac{1}{N_{\mathrm{c}}} D_{ii}(f) D_{jj}(f).
\end{equation}

Next, we define logarithmic frequency bins such that each decade in frequency contains approximately 30 bins. For each bin indexed by $k$, the binned frequency $f_{ij}^{(k)}$ and the corresponding data $\mathcal{D}_{ij}^{(k)}$ are computed using inverse-variance weighting~\cite{Caprini:2019pxz}
\begin{equation}
f_{ij}^{(k)} = \frac{\sum_{f \in \text{bin}_k} w_{ij}(f)  f}{\sum_{f \in \text{bin}_k} w_{ij}(f)}, \quad
\mathcal{D}_{ij}^{(k)} = \frac{\sum_{f \in \text{bin}_k} w_{ij}(f) \bar{D}_{ij}(f)}{\sum_{f \in \text{bin}_k} w_{ij}(f)}.
\end{equation}
where $w_{ij}(f) = 1/\sigma_{ij}^2(f)$ is the inverse-variance weight, optimizing sensitivity by emphasizing low-noise frequency bins. Additionally, the number of data points within each bin is recorded as $N_{ij}^{(k)}$.\\

\textbf{Step 3: Data likelihood}\\

As a starting point for constructing the data likelihood function, we adopt a simple Gaussian likelihood assumption~\cite{Flauger:2020qyi}
\begin{equation}
\ln \mathcal{L}_G(D \mid \vec{\theta}, \vec{n}) = -\frac{N_c}{2} \sum_{i, j} \sum_k N_{ij}^{(k)} \left[ \frac{\mathcal{D}_{ij}^{\text{th}}\left(f_{ij}^{(k)}, \vec{\theta}, \vec{n}\right) - \mathcal{D}_{ij}^{(k)}}{\mathcal{D}_{ij}^{\text{th}}\left(f_{ij}^{(k)}, \vec{\theta}, \vec{n}\right)} \right]^2
\end{equation}
The indices $i, j$ run over different channel combinations, while the index $k$ spans the binned frequency points. The quantity $\mathcal{D}_{ij}^{\mathrm{th},(k)}$ denotes the theoretical prediction evaluated at the binned frequency $f_{ij}^{(k)}$. In the XYZ basis, the three channels are correlated, with non-zero off-diagonal elements in the data covariance. By transforming to the AET basis, we diagonalize the data streams. In this AET basis, the likelihood function simplifies to a sum over independent channel contributions, facilitating the separation of signal and noise components.

Although the Gaussian likelihood function is commonly adopted, it may introduce systematic bias due to mild non-Gaussianity in the full likelihood of non-averaged data. To mitigate this bias, we also incorporate a log-normal likelihood function~\cite{Bond:1998qg,Sievers:2002tq}
\begin{equation}
\ln \mathcal{L}_{\text{LN}}(D \mid \vec{\theta}, \vec{n}) = -\frac{N_c}{2} \sum_{i, j} \sum_k N_{ij}^{(k)} \left[ \ln \left( \frac{\mathcal{D}_{ij}^{\text{th}}\left(f_{ij}^{(k)}, \vec{\theta}, \vec{n}\right)}{\mathcal{D}_{ij}^{(k)}} \right) \right]^2.
\end{equation}

We define our final likelihood function as~\cite{WMAP:2003pyh}
\begin{equation}
\ln \mathcal{L} = \frac{1}{3} \ln \mathcal{L}_G + \frac{2}{3} \ln \mathcal{L}_{\text{LN}}.
\end{equation}

Table~\ref{tab:tqp} summarizes the TianQin basic parameters~\cite{TianQin:2015yph}.
\begin{table}[h!]
	\begin{center}
	\centering
	\caption{TianQin basic parameters.}
	\begin{tabular}{ccc}
		\hline\hline
		Parameter & Description & Value \\
        \hline\hline
		$L$ & Arm length & $\sqrt{3} \times 10^8$ m \\
		$T_{\mathrm{obs}}$ & Total effective observation time & 4 years \\
        $S_p(f)$ & Position noise & $1 \times 10^{-24}$ m$^2$/Hz\\
		$S_a(f)$ & Acceleration noise &  $1 \times 10^{-30}$ m$^2$ s$^{-4}$/Hz\\
		$f_{\mathrm{range}}$ & Frequency range & $[10^{-4}, 1]$ Hz \\
        \hline\hline
	\end{tabular}\label{tab:tqp}
	\end{center}
\end{table}

\subsection{Framework for Statistical Parameter Estimation}

To extract geometric parameters from simulated data, we employ two complementary statistical inference methods: Fisher matrix analysis~\cite{Gowling:2021gcy} and \texttt{PolyChord}~\cite{handley2015polychord,Handley:2015fda}. This dual-strategy approach enables both rapid assessment of the detector's overall sensitivity across a broad parameter space and detailed reconstruction of specific signal features. The Fisher matrix provides a computationally efficient estimate of expected uncertainties, while \texttt{PolyChord} offers a Bayesian framework for exploring complex likelihood and quantifying posterior distributions.

\textbf{Fisher matrix analysis}:
Fisher matrix analysis is a powerful tool for rapidly estimating the precision of parameter measurements. It provides a theoretical lower bound on parameter uncertainties by evaluating the curvature of the likelihood function in the vicinity of the true parameter values. For parameter estimation in the context of a SGWB, the Fisher matrix is defined as:
\begin{equation}
F_{ij} = T_{\text{obs}} \sum_{I=\mathrm{A, E, T}} \int_{f_{\min}}^{f_{\max}} \frac{1}{\Omega_{I}^2(f)} \frac{\partial \Omega_{I}}{\partial \theta_i} \frac{\partial \Omega_{I}}{\partial \theta_j} \, df
\end{equation}
where $\Omega_I h^2 = \Omega_{\mathrm{GW}}^{\mathrm{SW}}h^2 + \Omega_{\mathrm{GW}}^{\mathrm{Ext}}h^2 +\Omega_{\mathrm{GW}}^{\mathrm{Gal}}h^2 +\Omega_{n, I}h^2 $ 
denotes the total energy spectral density of both signal and noise in channel $I={A,E,T}$, and $\vec{\theta}$ represents the vector of parameters to be estimated.  The covariance matrix of the parameters is defined as the inverse of the Fisher matrix
\begin{equation}
C_{ij} = (F^{-1})_{ij}.
\end{equation}

The diagonal elements of the covariance matrix, $C_{ii}$, represent the variance of the $i$-th parameter after marginalizing over all other parameters, and thus quantify its uncertainty. The off-diagonal elements, $C_{ij}$ ($i \neq j$), characterize the covariance between different parameters. The Fisher matrix offers high computational efficiency, but it relies on the assumption of a perfectly Gaussian likelihood and a locally linear parameter space. As a result, it is only valid in scenarios with high signal-to-noise ratios (SNRs) and weak parameter degeneracies. In realistic observational settings, non-Gaussian noise, foreground contamination, and strong degeneracies may render Fisher-based forecasts overly optimistic.

\textbf{Bayesian Inference and Nested Sampling (\texttt{PolyChord})}:
While the Fisher matrix provides a fast and efficient estimate of parameter uncertainties, it is inherently limited to local linearized approximations. As such, it cannot capture non-Gaussian features or multimodal structures in the parameter space. To obtain the full posterior probability distribution and properly account for complex parameter degeneracies, we adopt a Bayesian inference framework. According to Bayes' theorem, the posterior probability of the geometric parameters is given by
\begin{equation}
	P(\vec{\theta}|D) =\frac{ \mathcal{L}(D|\vec{\theta}) \pi(\vec{\theta})}{Z},
\end{equation}
where $\mathcal{L}(D|\vec{\theta})$ is the likelihood function, $\pi(\vec{\theta})$ denotes the prior probability distribution, and $Z = \int \mathcal{L}(D | \vec{\theta}) \pi(\vec{\theta}) \, d\vec{\theta}$ is the Bayesian evidence (also known as the marginal likelihood), which serves as a key quantity for model comparison.

We employ \texttt{PolyChord}, an advanced nested sampling algorithm, to explore the posterior probability distribution. Compared to traditional Markov Chain Monte Carlo (MCMC) methods, \texttt{PolyChord} demonstrates superior performance in handling multimodal distributions and complex parameter degeneracies, while simultaneously computing the Bayesian evidence. Although the computational cost of \texttt{PolyChord} is significantly higher than that of the Fisher matrix approach, it yields the full posterior distribution of the parameters. Its output includes posterior samples, marginalized distributions, and the correlation matrix among parameters, providing comprehensive insight into the strength and structure of parameter constraints.

The complete parameter reconstruction pipeline, from data generation to model parameter, is schematically presented in Fig.~\ref{fig:flow}. We first generate simulated data by combining TDI channel noise, astrophysical foregrounds, and GW signals from the dimension-six model. The simulated data undergoes parameter estimation through two complementary methods: Fisher matrix analysis for rapid parameter estimation and Bayesian nested sampling (\texttt{PolyChord}) for handling high-dimensional, multi-modal posterior distributions. These methods yield posterior distributions of geometric parameters $\left(\Omega_2, f_1, f_2\right)$. Finally, we employ an ensemble of machine-learning algorithms-Gaussian Process Regression, Random Forest, Gradient Boosting Trees, and Multi-Layer Perceptron to map the geometric parameters to the model parameter $\Lambda$.

\section{Reconstruction Results in the Dimension-Six Model with TianQin}\label{sec:res_tq}

In this section, we carry out the final step illustrated in Fig.~\ref{fig:flow}, namely the evaluation of reconstruction capabilities of the TianQin detector for EWPT within the framework of the dimension-six SMEFT. We first extract the key phase transition parameters characterizing the SFOPT for a given model parameter $\Lambda$, namely the percolation temperature $T_p$, the phase transition strength $\alpha$, and the normalized inverse duration parameter $\beta/H_p$. These quantities are used as inputs to model the resulting stochastic GW signal. We then evaluate the detector sensitivities to both the geometric parameters of the GW spectrum and the underlying physical parameters of the dimension-six model.

\subsection{Phase transition parameters in the dimension-six model}
\begin{table}[h!]
\begin{center}
\centering
\caption{Phase transition parameters for three benchmark points in the dimension-six model.}
\begin{tabular}{ccccc}
\hline\hline
 & $\Lambda$ [GeV] & $T_p$ [GeV] & $\alpha$ & $ \beta/H_p$ \\
\hline\hline
$\mathrm{BP_1}$ & 548.31 & 36.25 & 0.55 & 4.42 \\
$\mathrm{BP_2}$ & 549.02 & 39.72 & 0.38 & 39.33 \\
$\mathrm{BP_3}$ & 550.16 & 43.32 & 0.27 & 81.39 \\
\hline\hline
\end{tabular}\label{tab:pt}
\end{center}
\end{table}
Table~\ref{tab:pt} summarizes three representative benchmark scenarios for the EWPT within the framework of the dimension-six model. The benchmark 
points are selected within the parameter range $\Lambda \in [548, 551]~\mathrm{GeV}$, where the phase transition strength is sufficiently large to produce GW signals detectable by TianQin or LISA. For each benchmark value of the model parameter $\Lambda$, we provide the corresponding phase transition parameters that govern the dynamics and determine the resulting GW signatures.
 
\subsection{Reconstruction Capabilities of TianQin}
\subsubsection{Reconstruction of the geometric parameters with TianQin}
\begin{figure}[h!]
	\centering
	\includegraphics[width=0.7\linewidth]{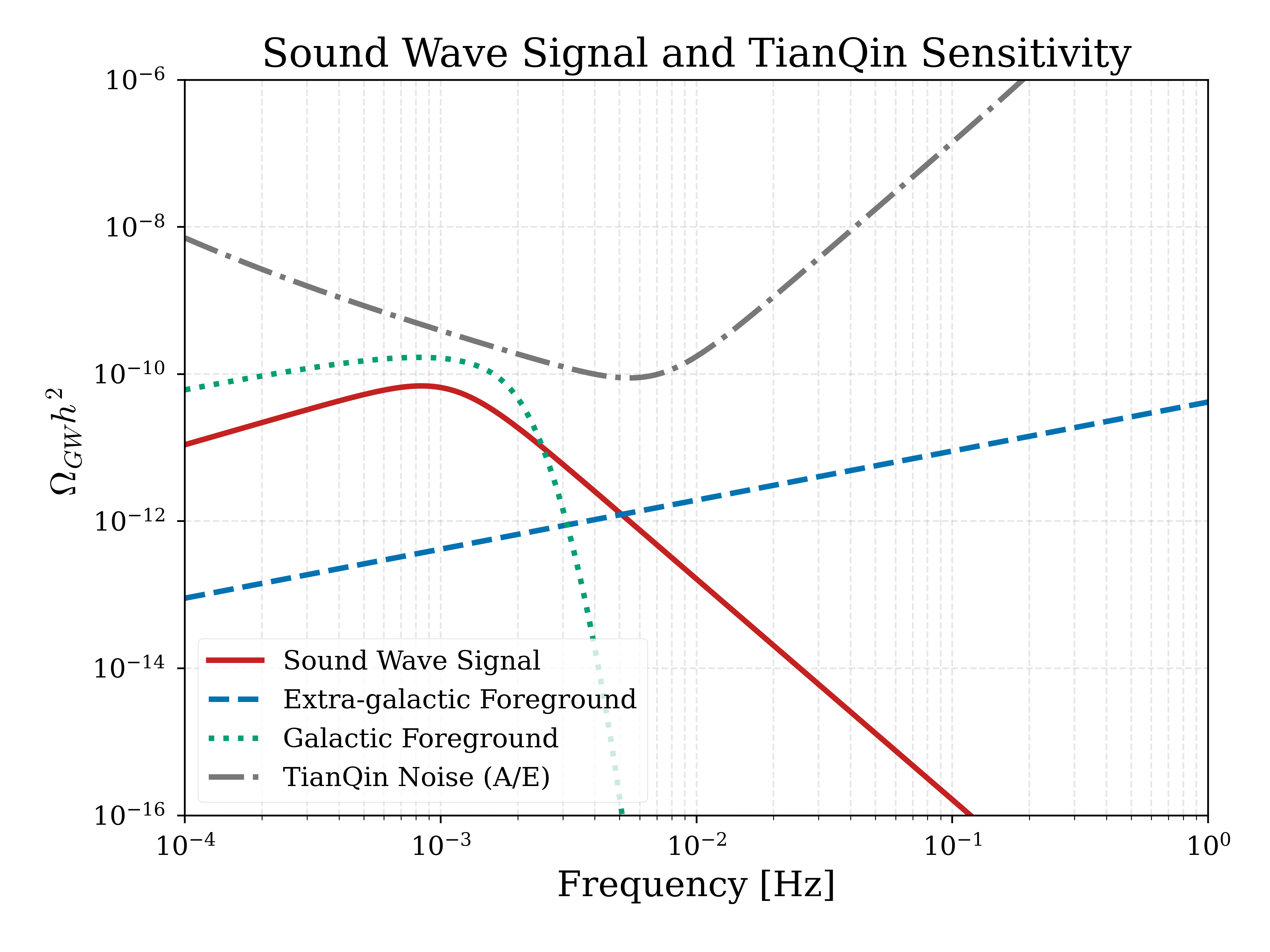}
	\caption{The GW spectrum for the $\mathrm{BP_1}$ of the dimension-six model, assuming a bubble wall velocity of $v_w=0.57$. The SW signal (solid red line) is shown against the sensitivity curve of the TianQin (gray dash-dotted line), along with foregrounds from galactic (dotted green) and extra-galactic (dashed blue) sources.}
	\label{fig:tqgwa}
\end{figure}

\begin{figure}
    \centering
    \includegraphics[width=0.7\linewidth]{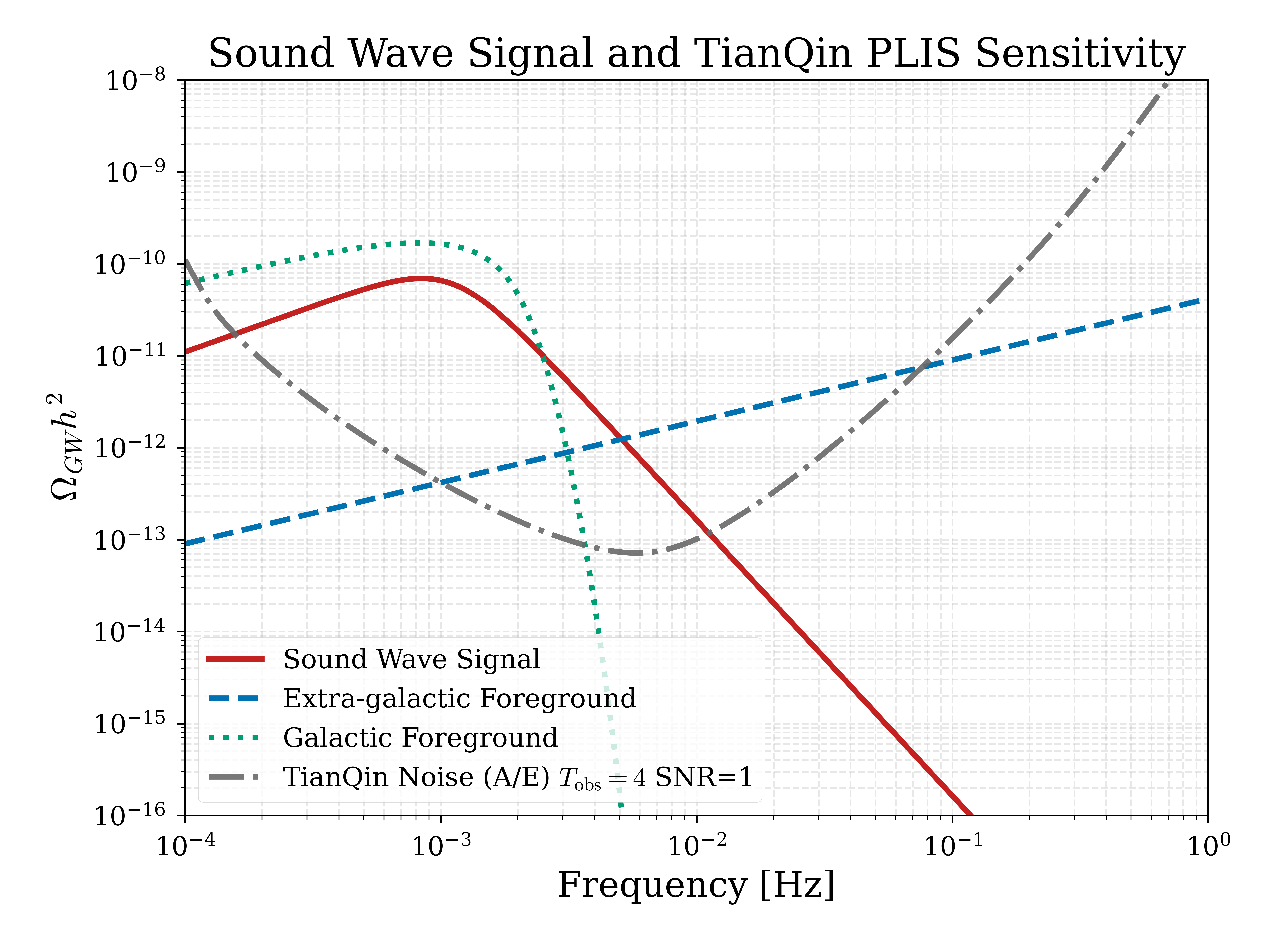}
    \caption{The power-law integrated sensitivity (PLIS) curve of TianQin for $\mathrm{SNR}=1$ with $T_{\mathrm{obs}} = 4~\mathrm{years}$ (gray dash-dotted line), compared with the SW signal from $\mathrm{BP_1}$ (solid red line). Although the signal lies below the sensitivity curve shown in Fig.~\ref{fig:tqgwa}, it exceeds the PLIS curve.}
    \label{figpls}
\end{figure}

\begin{figure}
    \centering
    \includegraphics[width=0.7\linewidth]{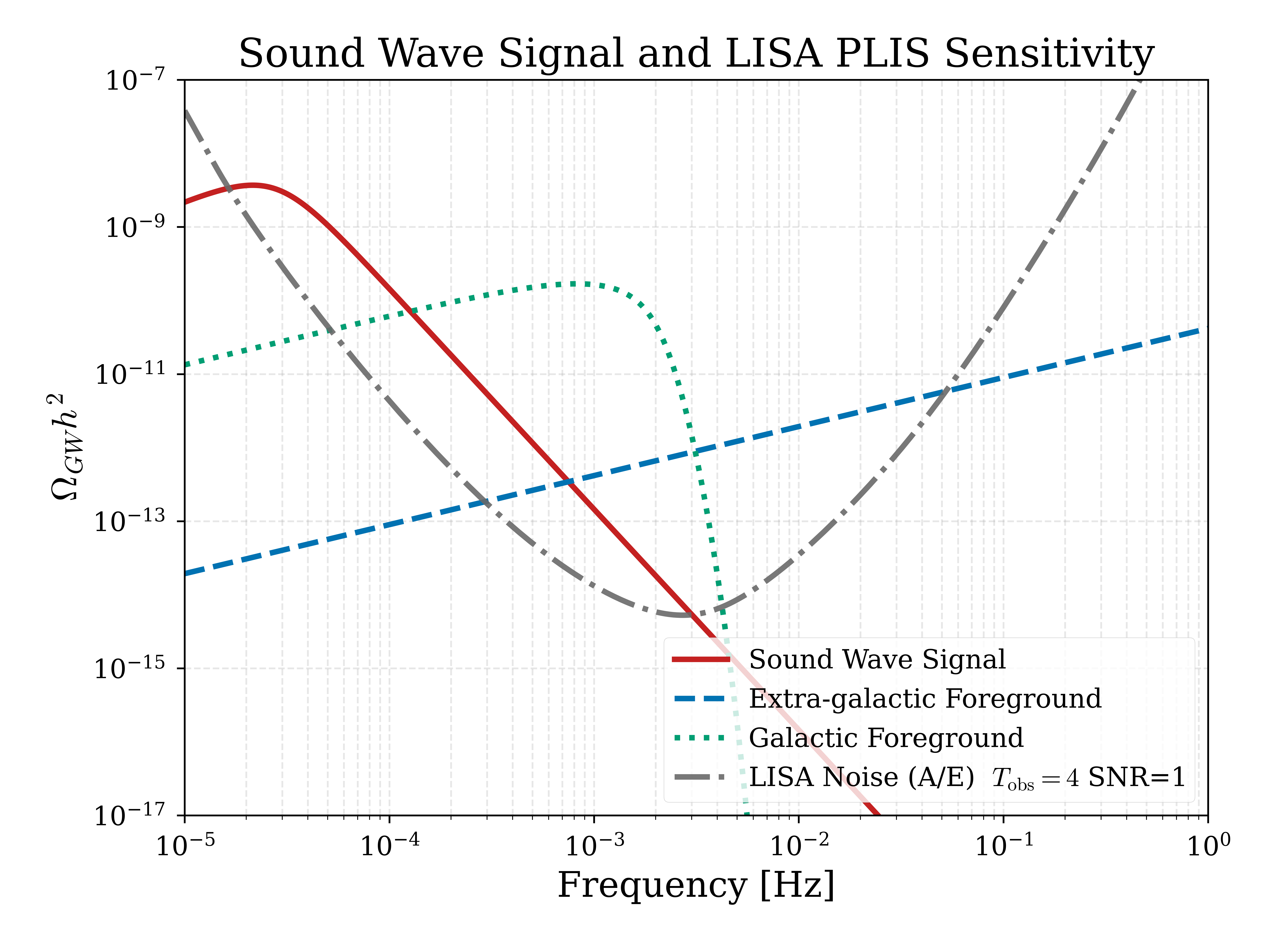}
    \caption{The PLIS curve of LISA for $\mathrm{SNR}=1$ with $T_{\mathrm{obs}} = 4~\mathrm{years}$ (gray dash-dotted line), compared with the SW signal from $\mathrm{BP_1}$ (solid red line).}
    \label{figpls2}
\end{figure}

First, we consider the benchmark point $\mathrm{BP_1}$, which is expected to produce a signal close to the maximum sensitivity of TianQin. Using the phase transition parameters for $\mathrm{BP_1}$ listed in Tab.~\ref{tab:pt}, we compute the corresponding GW spectrum from the SW component, modeled by the DBPL spectrum, as shown in Fig.~\ref{fig:tqgwa}. 
The predicted GW signal is compared with the TianQin sensitivity curve in Fig.~\ref{fig:tqgwa}. 
As shown in the figure, the signal amplitude lies below the detector sensitivity curve across all frequencies. However, this does not preclude detection. The detectability of the GW signal can be estimated through its SNR, which may be enhanced by extending the observation time sufficiently.  With $T_{\mathrm{obs}} = 4~\mathrm{years}$, the signal can significantly exceed the PLIS curve as shown in Fig.~\ref{figpls} and Fig.~\ref{figpls2}.

More specifically, the detectability of the signal depends on whether its SNR exceeds the threshold required by the detector. Since the SNR is determined through time integration, extending the observation duration enhances the likelihood of detection.

Second, we analyze the capability of the TianQin detector to reconstruct the geometric parameters of the GW signal using both the Fisher matrix approach and the nested sampling algorithm implemented with \texttt{PolyChord}. We adopt benchmark point $\mathrm{BP_1}$ with bubble wall velocity $v_w = 0.57$ to reconstruct three key geometric parameters of the DBPL template: $\log_{10}(h^2\Omega_2)$, $\log_{10}(f_2/\text{Hz})$, and $\log_{10}(f_2/f_1)$. In principle, $v_w$ can be determined by solving the Boltzmann-fluid equations governing the interaction between the bubble wall and the surrounding plasma~\cite{Ekstedt:2024fyq,Laurent:2022jrs}. Utilizing numerical simulations and theoretical methods allows for a more accurate estimation of the bubble wall velocity.  However, this calculation constitutes an independent and involved analysis that lies beyond the primary scope of this work, which focuses on the reconstruction pipeline from GW observables to model parameters. We therefore treat $v_w$ as a fixed external input. This choice of $v_w$ is motivated by its ability to produce a GW signal that matches the peak sensitivity of TianQin, allowing high-precision reconstruction of the spectral features. In addition to signal reconstruction, we also present the posterior distributions of the foreground amplitudes from Galactic and extragalactic sources.
\begin{figure}[h!]
	\centering
\includegraphics[width=1\linewidth]{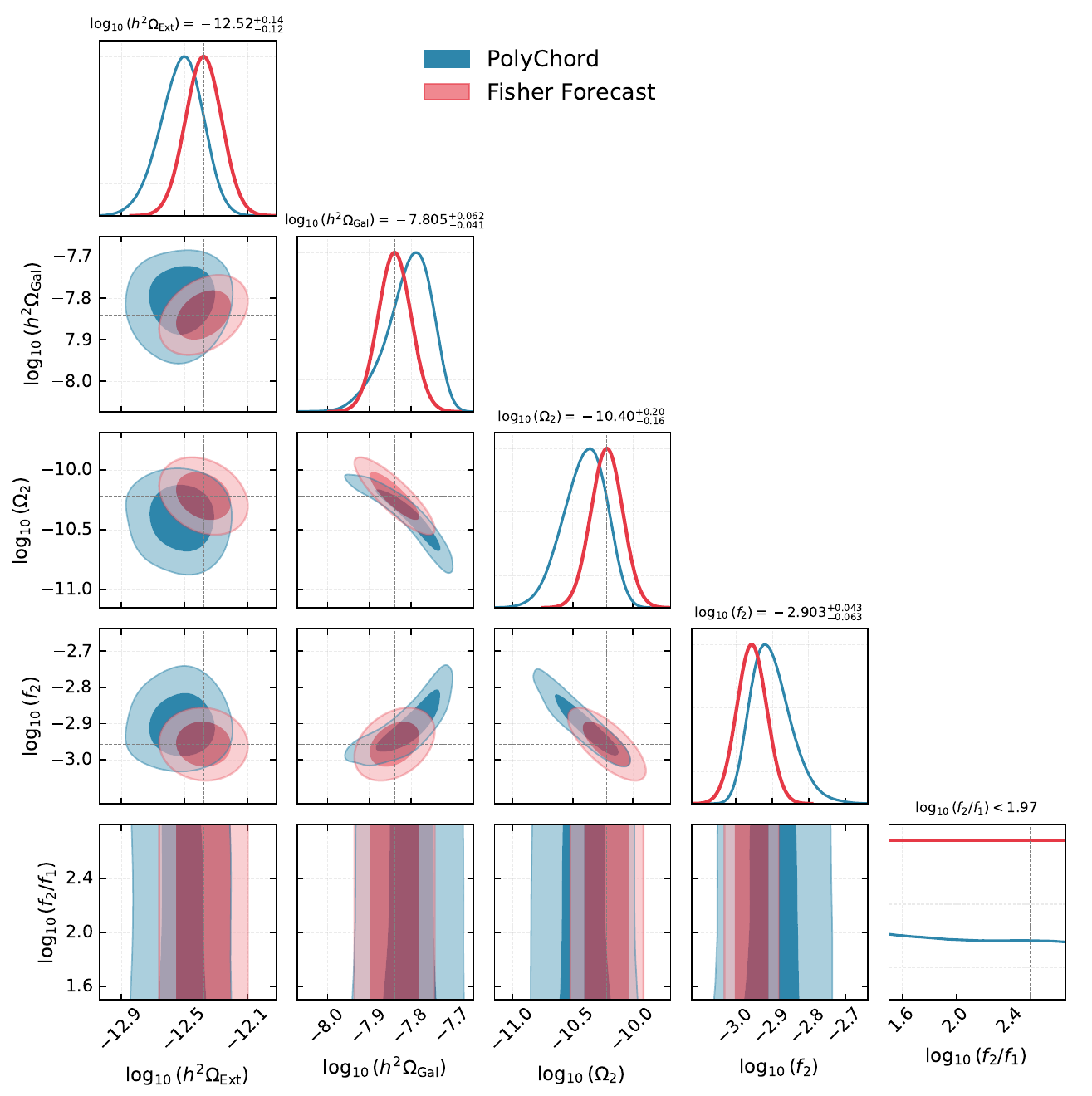}
	\caption{ Triangle plot comparing the geometric parameter estimation from a Fisher forecast (red contours) and a \texttt{PolyChord}  (blue contours). The analysis is performed on a simulated GW signal $\mathrm{BP_1}$ with a fiducial amplitude $\Omega_2 = 6.09 \times 10^{-11}$, break frequency $f_2 = 1.10~\mathrm{mHz}$, and frequency ratio $f_2/f_1 \approx 348$ described by a DBPL template. The simulation includes instrumental noise for the TianQin detector and astrophysical foregrounds.}
	\label{fig:tqtri}
\end{figure}

The results are illustrated in the triangle plot shown in Fig.~\ref{fig:tqtri}. In this type of visualization, the diagonal panels display the one-dimensional marginalized posterior distributions for each parameter, while the off-diagonal panels present the two-dimensional joint posterior contours, with the dark shaded regions representing the $68\%$  credible interval (CI) and the light shaded regions representing the $95\%$ CI. We reconstruct a dimension-six signal from the SW component with a fiducial amplitude $\Omega_2 = 6.09 \times 10^{-11}$, break frequency $f_2 = 1.10~\mathrm{mHz}$, and frequency ratio $f_2/f_1 \approx 348$ described by a DBPL template. For the reconstruction of the geometric parameters, compact and approximately elliptical confidence contours are expected, indicating that TianQin can determine the spectral shape of the GW signal with high precision. Indeed, the figure shows such elliptical contours, confirming TianQin's capability for accurate spectral reconstruction. This figure also demonstrates good consistency between the Fisher matrix and \texttt{PolyChord} methods, although the Fisher-based constraints appear slightly tighter. For well-constrained parameters, such as the signal amplitude $\log_{10}(\Omega_2)$, TianQin achieves a relative error (at $68\%$ CI) of approximately $29.19\%$ for the amplitude. A notable negative correlation is observed between $\log_{10}(\Omega_2)$ and $\log_{10}(f_2)$, as evidenced by the upward-left tilt of the corresponding contour. This reflects a compensatory relationship, where a larger amplitude can be partially offset by a lower peak frequency. In contrast, the one-dimensional posterior distribution of $\log_{10}(f_2 / f_1)$ is nearly flat and does not form a closed constraint region. This suggests that, at the current SNR, the data are insensitive to the low-frequency portion of the signal, making it difficult to accurately determine $f_1$. Finally, the extragalactic foreground parameter $\log_{10}(h^2 \Omega_{\mathrm{Ext}})$ exhibits nearly circular contours with respect to other parameters, indicating weak correlations and minimal degeneracy.
\begin{figure}[h!]
	\centering
	\includegraphics[width=0.80\linewidth]{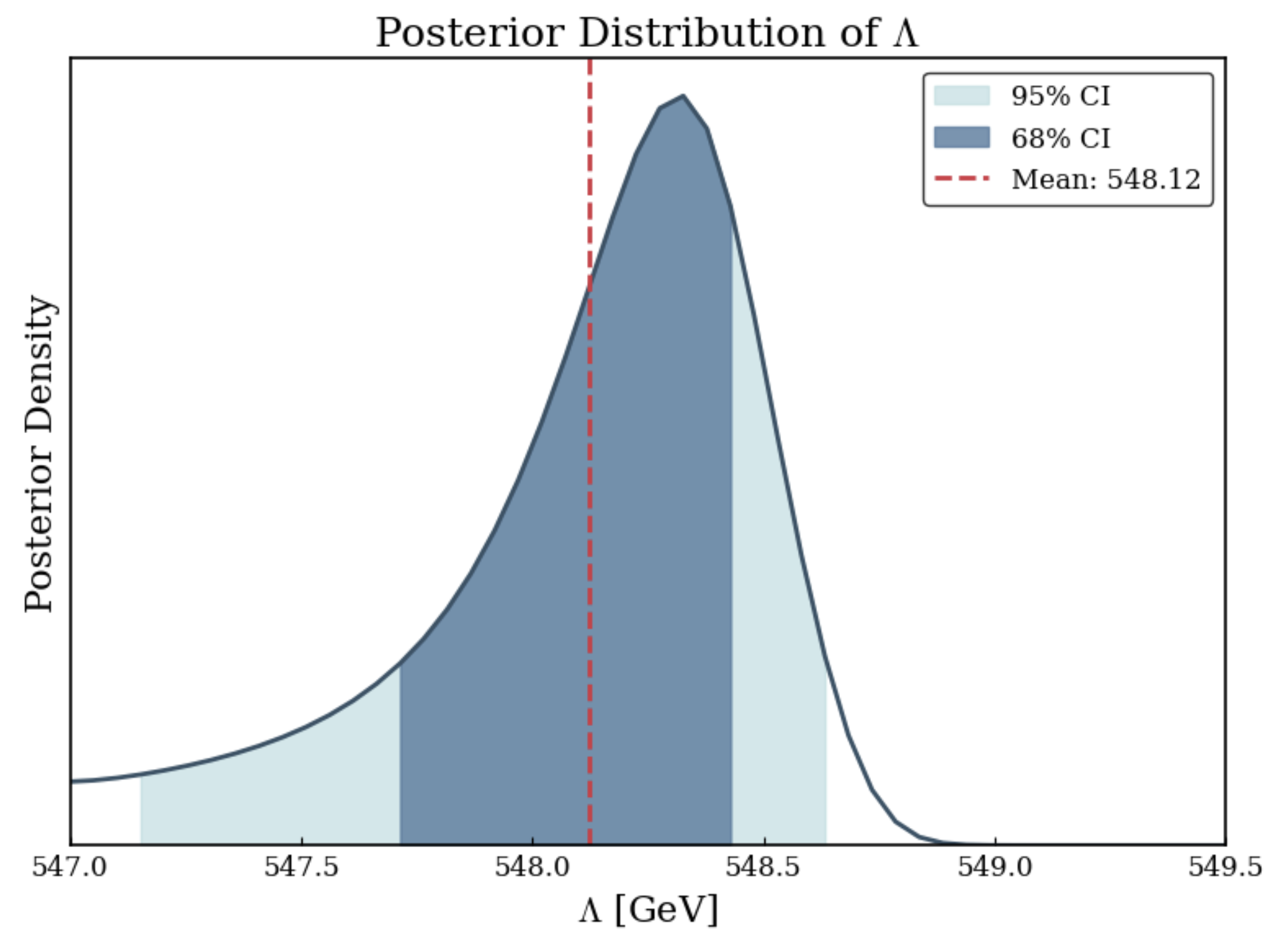}
	\caption{The posterior probability distribution for the reconstruction of the parameter $\Lambda$ for $\mathrm{BP_1}$, based on a simulated analysis for the TianQin detector.  The mean of the posterior is indicated by the red dashed line at $\Lambda=548.12~\mathrm{GeV}$ with a standard deviation of $0.36~\mathrm{GeV}$. The shaded regions represent the 68\% (darker blue) and 95\% (lighter blue) CIs.}
	\label{fig:lamtq}
\end{figure}
\subsubsection{Reconstruction of model Parameter with TianQin}
Finally, we reconstruct the model parameter $\Lambda$ from the simulated SGWB signal for  $\mathrm{BP_1}$. The posterior distribution yields a mean value of $\Lambda = 548.12~\mathrm{GeV}$ with a standard deviation of $0.36~\mathrm{GeV}$. The $68\%$ CI  spans $[547.74, 548.41]~\mathrm{GeV}$, while the $95\%$ CI  extends to $[547.12, 548.65]~\mathrm{GeV}$.  The injected true value $\Lambda_{\rm true} = 548.31~\mathrm{GeV}$ lies within the $68\%$ CI, demonstrating excellent consistency between the reconstructed and injected parameters. This result indicates that TianQin can constrain the cutoff scale $\Lambda$ with sub-percent accuracy for signals of this amplitude.

\section{Similar discussions for LISA}\label{sec:res_lisa}
The parameter reconstruction framework for LISA follows the same pipeline as TianQin, including the same foreground modeling (extragalactic and Galactic foregrounds), data generation procedure, and Bayesian statistical framework. The key differences are LISA's sensitivity band of $[10^{-5}, 1]~\mathrm{Hz}$, its noise model, and response transfer functions. Given these detector-specific differences,  we evaluate the reconstruction capabilities of the LISA detector for EWPT within the framework of the dimension-six effective field theory in this section.

\subsection{Noise model of LISA}

In the context of GW detection with LISA, the dominant sources of statistical uncertainty arise from two primary factors: perturbations to the free-fall motion of the TMs and uncertainties in the relative position measurements between the TM. The total noise is also composed of contributions from both the TM noise and the OMS noise. The corresponding PSD can be expressed as~\cite{Caprini:2019pxz,Flauger:2020qyi}
\begin{equation}
\begin{aligned}
S_{\mathrm{TM}}(f) & =A_{\mathrm{TM}}^2\left(1+\left(\frac{0.4 \mathrm{mHz}}{f}\right)^2\right)\left(1+\left(\frac{f}{8 \mathrm{mHz}}\right)^4\right)\left(\frac{1}{2 \pi f c}\right)^2\left(\frac{\mathrm{fm}^2}{\mathrm{~s}^3}\right), \\
S_{\mathrm{OMS}}(f) & =P_{\mathrm{OMS}}^2\left(1+\left(\frac{2 \times 10^{-3} \mathrm{mHz}}{f}\right)^4\right)\left(\frac{2 \pi f}{c}\right)^2\left(\frac{\mathrm{pm}^2}{\mathrm{~Hz}}\right).
\end{aligned}
\end{equation}
The term $S_{\mathrm{TM}}$ characterizes noise induced by deviations from ideal free-fall motion, while $S_{\mathrm{OMS}}$ represents noise arising from optical path-length measurement errors. In the simulations, we inject noise using parameters $A_{\mathrm{TM}} = 3$ and $P_{\mathrm{OMS}} = 15$.

To compute the LISA noise in the XYZ basis, we adopt the following assumptions: for the interferometric measurement system, the optical path-length noise in all three arms is assumed to have identical PSD, to be stationary in time, and mutually uncorrelated. For the TM acceleration noise, the disturbances acting on the six TMs are assumed to be isotropic, stationary, and statistically independent across different masses. Under these assumptions, the noise power spectral densities of the XYZ channels can be expressed as linear combinations of the TM and interferometric measurement noises. We then transform to the AET basis, which diagonalizes the covariance matrix of both signal and noise. The AET channels are constructed as specific linear combinations of the XYZ channels, yielding~\cite{Flauger:2020qyi}
\begin{equation}
\begin{aligned}
&\begin{aligned}
N_{\mathrm{A}}(f)=N_{\mathrm{E}}(f)= & \\
=8 \sin ^2\left(\frac{2 \pi f L}{c}\right) & \left\{4\left[1+\cos \left(\frac{2 \pi f L}{c}\right)+\cos ^2\left(\frac{2 \pi f L}{c}\right)\right] S_{\mathrm{TM}}(f)+\right. \\
& \left.+\left[2+\cos \left(\frac{2 \pi f L}{c}\right)\right] S_{\mathrm{OMS}}(f)\right\},
\end{aligned}\\
&\begin{aligned}
N_{\mathrm{T}}(f)=16 \sin ^2\left(\frac{2 \pi f L}{c}\right) & \left\{2\left[1-\cos \left(\frac{2 \pi f L}{c}\right)\right]^2 S_{\mathrm{TM}}(f)+\right. \\
+ & {\left.\left[1-\cos \left(\frac{2 \pi f L}{c}\right)\right] S_{\mathrm{OMS}}(f)\right\} }.
\end{aligned}
\end{aligned}
\end{equation}

\begin{figure}[h!]
	\centering
\includegraphics[width=0.7\linewidth]{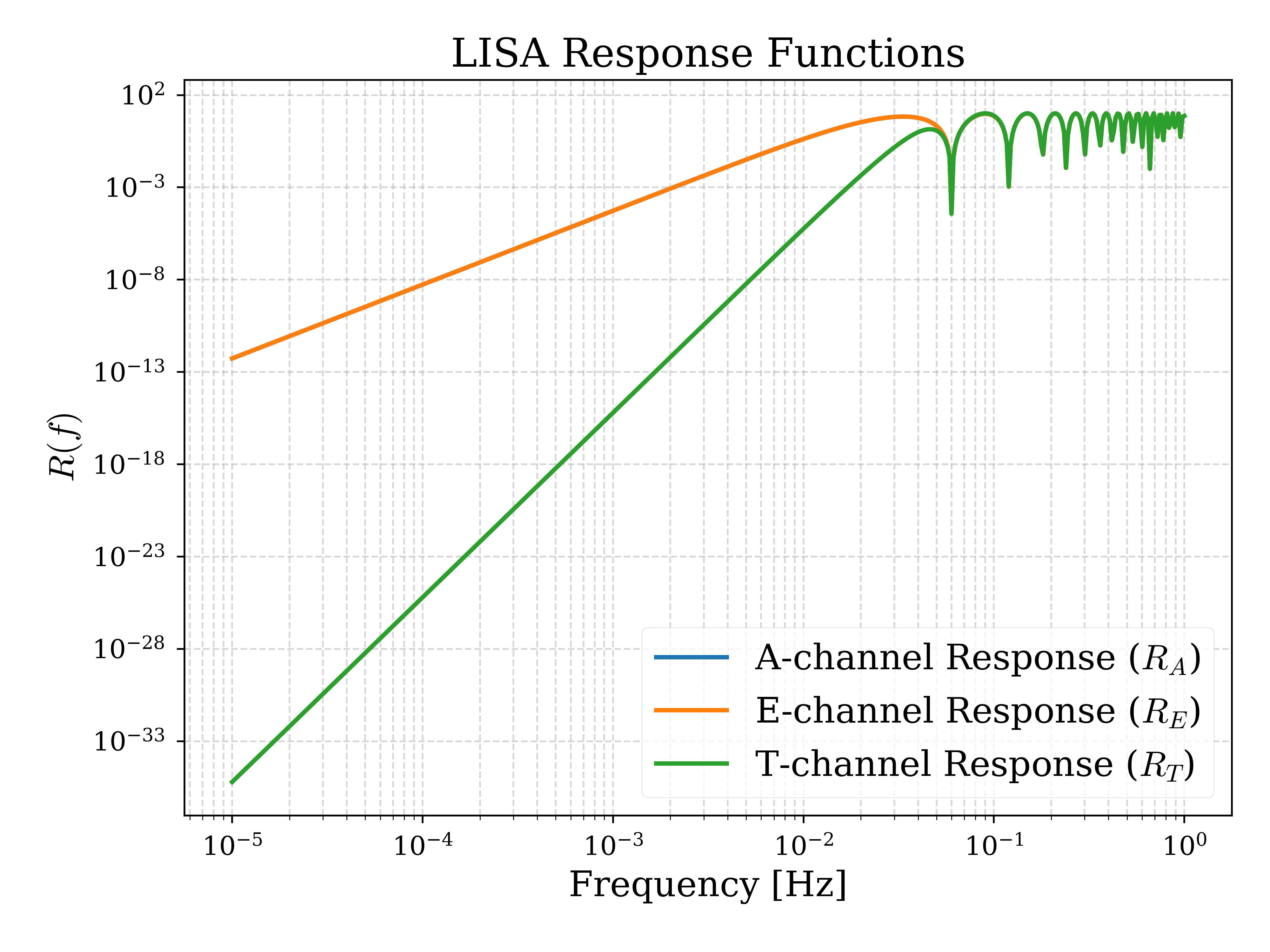}
	\caption{LISA response function in the AET Basis.}
	\label{fig:lisares}
\end{figure}

Figure~\ref{fig:lisares} illustrates the frequency-dependent response functions of LISA's three AET channels. The A-channel ($R_A$, blue) and E-channel ($R_E$, orange). The T-channel ($R_T$, green) shows minimal response at low frequencies and becomes significant only at higher frequencies. The mathematical form of the response function is given by~\cite{Flauger:2020qyi}
\begin{equation}
\mathcal{R}_{i j}(f)=16 \sin ^2\left(\frac{2 \pi f L}{c}\right)\left(\frac{2 \pi f L}{c}\right)^2 \tilde{R}_{i j}(f),
\end{equation}
where the first factor encodes the TDI combination characteristics, the ($2 \pi f L / c$) term reflects the frequency-measurement nature of the detector, and $\tilde{R}_{i j}$ represents the geometric configuration factor. For the AA and EE channels, the geometric factor can be approximated as
\begin{equation}
\tilde{R}_{\mathrm{A}}(f)=\tilde{R}_{\mathrm{E}}(f) \approx \frac{9}{20} \frac{1}{1+0.7\left(\frac{2 \pi f L}{c}\right)^2},
\end{equation}
while the T-channel exhibits higher-order frequency dependence
\begin{equation}
\tilde{R}_{\mathrm{T}}(f) \approx \frac{9}{20} \frac{\left(\frac{2 \pi f L}{c}\right)^6}{1.8 \times 10^3+0.7\left(\frac{2 \pi f L}{c}\right)^8}.
\end{equation}

\begin{figure}[h!]
	\centering
	\includegraphics[width=0.7\linewidth]{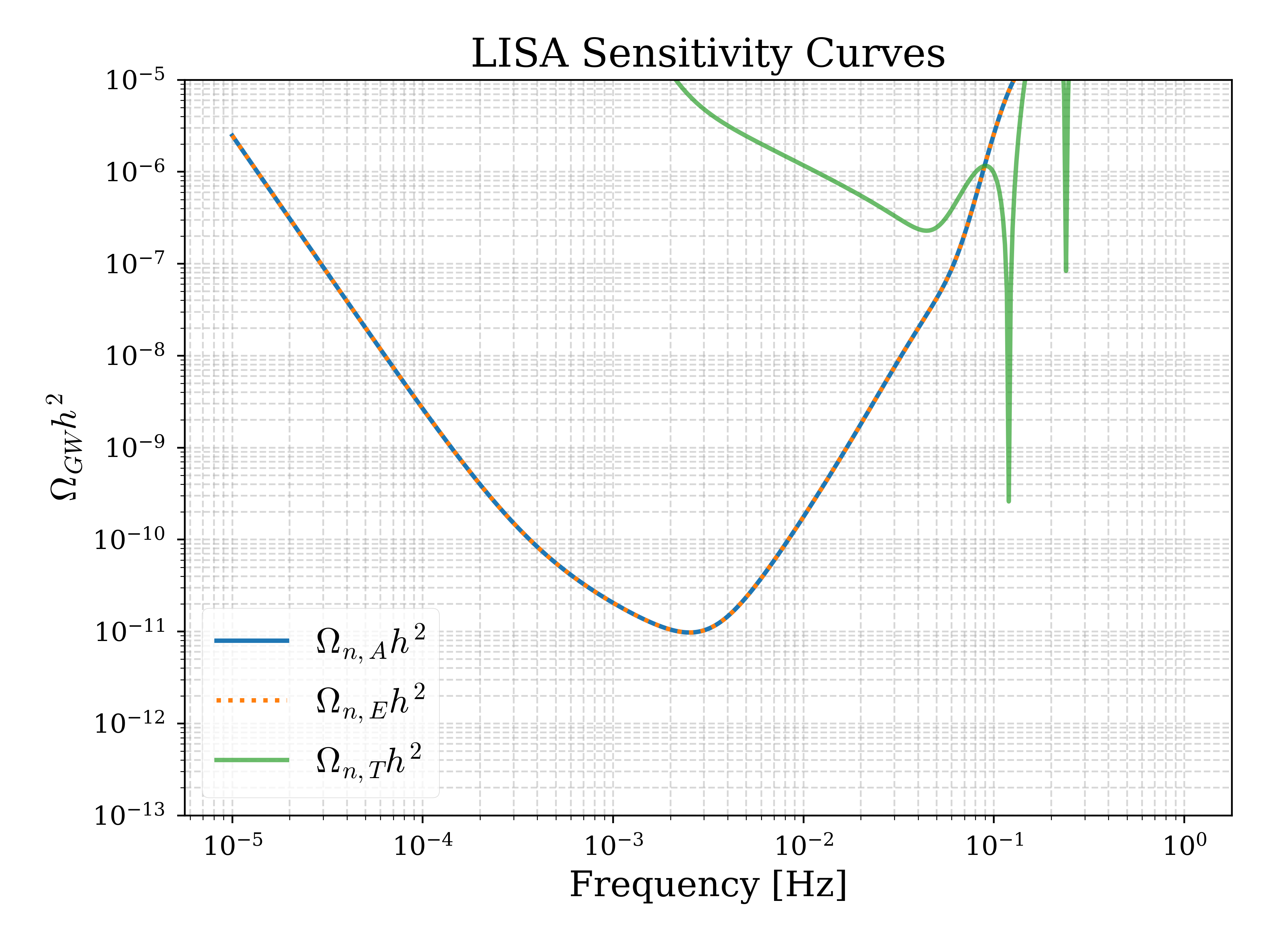}
	\caption{ LISA's sensitivity in the AET Basis.}
	\label{fig:lisasen}
\end{figure}

Figure~\ref{fig:lisasen} illustrates the LISA sensitivity in the AET channel basis, expressed in terms of the GW energy density parameter $\Omega_{\mathrm{GW}} h^2$. The sensitivity curves for the A channel ($\Omega_{n, A} h^2$, solid blue line) and the E channel ($\Omega_{n, E} h^2$, dashed orange line) are nearly identical. In contrast, the T channel ($\Omega_{n, T} h^2$, solid green line) exhibits significantly reduced sensitivity at low frequencies and displays oscillatory features at high frequencies due to light travel time effects. The energy density sensitivity is the same as Eq.~\eqref{eq:ed}.

\subsection{Reconstruction of the geometric parameters with LISA}

\begin{figure}
	\centering
	\includegraphics[width=0.7\linewidth]{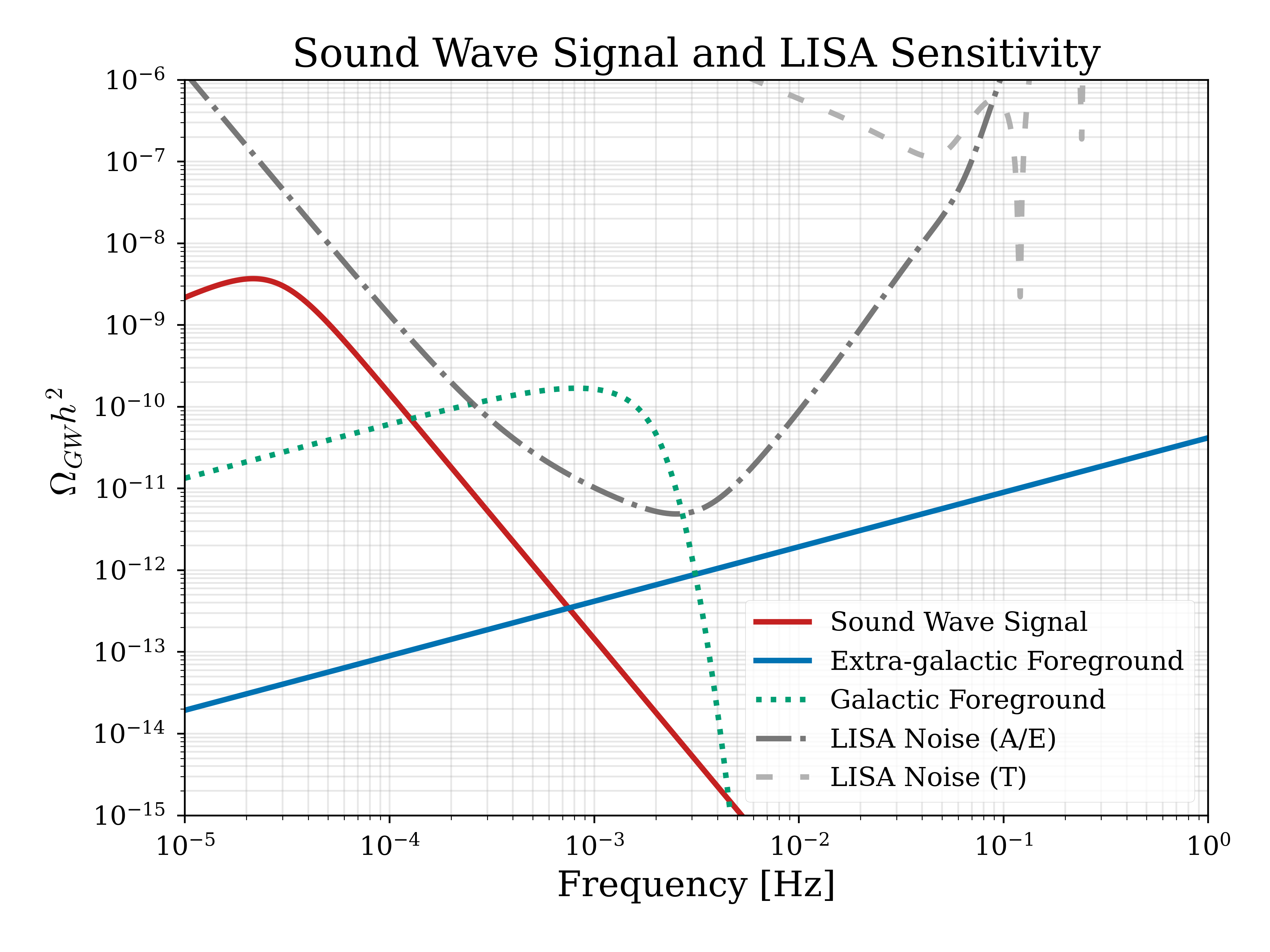}
	\caption{The GW spectrum for the $\mathrm{BP_1}$ of the dimension-six model with LISA, assuming a bubble wall velocity of $v_w=0.74$. The SW signal (solid red line) is shown against the sensitivity curve of the LISA (gray dash-dotted line), along with foregrounds from galactic (dotted green) and extra-galactic (dashed blue) sources.}
	\label{fig:lisagwsp}
\end{figure}

Using the same set of parameters listed in Tab.~\ref{tab:pt}, Fig.~\ref{fig:lisagwsp} shows the response of the LISA detector to the GW signal. To maximize the number of observable signals, we adopt a bubble wall velocity of $v_w = 0.74$ in our analysis. Under this parameter configuration, LISA is sensitive not only to the GW signal generated by $\mathrm{BP_1}$ but also capable of reconstructing signals from $\mathrm{BP_2}$ and $\mathrm{BP_3}$, thereby covering all three benchmark scenarios. Unlike TianQin, LISA exhibits peak sensitivity around $10^{-3}~\mathrm{~Hz}$, which better matches the GW signal frequency predicted by the dimension-six model.

\begin{figure}[h!]
	\centering
	\includegraphics[width=1\linewidth]{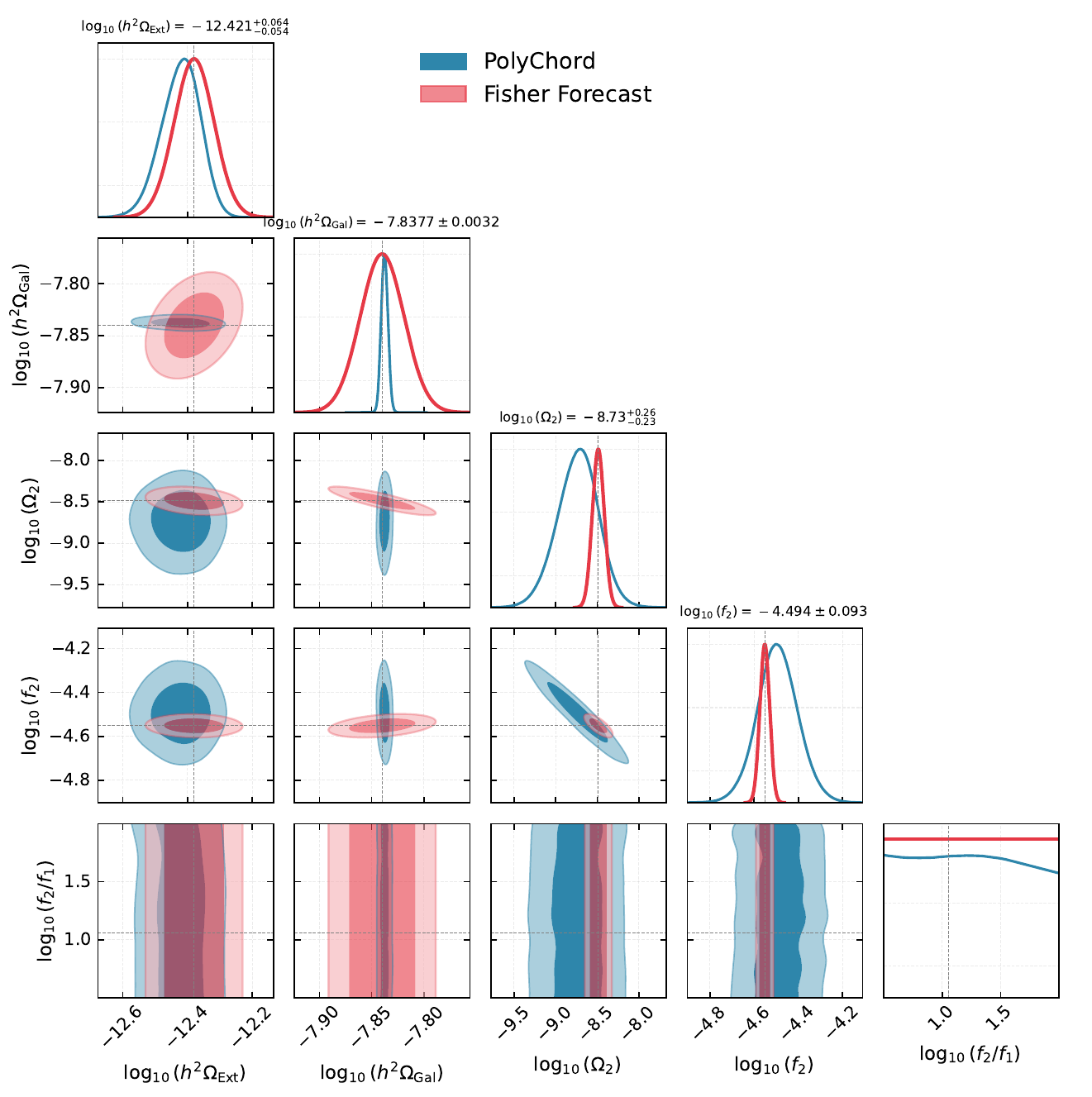}
	\caption{Triangle plot comparing the geometric parameter estimation from a Fisher forecast (red contours) and a \texttt{PolyChord}  (blue contours). The analysis is performed on a simulated GW signal $\mathrm{BP_1}$ with fiducial amplitude $\Omega_2 = 3.26 \times 10^{-9}$, break frequency $f_2 = 28.07~\mu\mathrm{Hz}$, and frequency ratio $f_2/f_1 \approx 11$.  The simulation includes instrumental noise for the LISA detector and astrophysical foregrounds.}
	\label{fig:lisatri}
\end{figure}

We also perform a geometric  parameter reconstruction analysis for $\mathrm{BP_1}$ with a fiducial amplitude $\Omega_2 = 3.26 \times 10^{-9}$, break frequency $f_2 = 28.07~\mu\mathrm{Hz}$, and frequency ratio $f_2/f_1 \approx 11$ shown in Fig.~\ref{fig:lisatri}. It is worth noting that the Fisher forecast (red) deviates from the \texttt{PolyChord} results (blue), in sharp contrast to the TianQin case. Since the signal lies above the foreground, nonlinear effects and parameter degeneracies become more pronounced, leading to a breakdown of the linear assumptions inherent in the Fisher approximation.

The following features can be observed in Fig.~\ref{fig:lisatri}. Both $\log_{10}(\Omega_2)$ and $\log_{10}(f_2)$ are well constrained, with the posterior distributions from \texttt{PolyChord} showing compact, elliptical contours. This indicates that LISA can accurately reconstruct these two key parameters. For the $\mathrm{BP_1}$ benchmark, TianQin demonstrates superior parameter reconstruction performance compared to LISA's analysis of similar phase transition scenarios. TianQin achieves relative uncertainties of approximately  $29.19\%$ for the signal amplitude, whereas LISA obtains a relative uncertainty of $\sim 30.15\%$  precision. This difference is primarily attributed to the signal frequencies: the $\mathrm{BP_1}$ signal peaks at $f_2 \sim 1.10$ mHz, which falls within TianQin's optimal sensitivity band, while LISA's analyzed signal at $f_2 \sim 28.07~\mu\mathrm{Hz}$ lies in a less sensitive region. However, LISA's analyzed signal has a higher amplitude than TianQin's, which compensates for its suboptimal frequency placement. This highlights LISA's advantage in low-frequency detection over TianQin. For $\mathrm{BP_2}$, LISA successfully constrains amplitudes with approximately $20.69\%$ precision. Overall, LISA exhibits stronger reconstruction capability for the dimension-six EWPT model across a broader parameter space.
The contour plot of $\log_{10}(\Omega_2)$ versus $\log_{10}(f_2)$ shows a clear tilt toward the upper left. This degeneracy pattern and posterior distribution of $\log_{10}(f_2 / f_1)$ are consistent with those found in the TianQin analysis. The Galactic foreground amplitude $\log_{10}(h^2 \Omega_{\mathrm{Gal}})$ is constrained with high precision, and its contour lines are nearly circular with respect to other signal parameters, indicating weak correlations. In contrast, the constraint on the extragalactic foreground $\log_{10}(h^2 \Omega_{\mathrm{Ext}})$ is comparatively weaker.

In summary, in regions where the precision of geometric parameter reconstruction is high, the uncertainties predicted by the Fisher analysis are in good agreement with the results obtained from \texttt{PolyChord}. In contrast, noticeable discrepancies appear in regions with lower reconstruction accuracy. This behavior is expected: the Fisher method, based on assumptions of local linearity and Gaussianity in parameter space, tends to yield overly optimistic constraints, resulting in tighter contours. By comparison, \texttt{PolyChord} performs full Bayesian sampling and captures the nonlinear structure of the parameter space, providing more reliable uncertainty estimates.

\subsection{Reconstruction of the model parameter with LISA}
\begin{figure}
	\centering
	\includegraphics[width=0.8\linewidth]{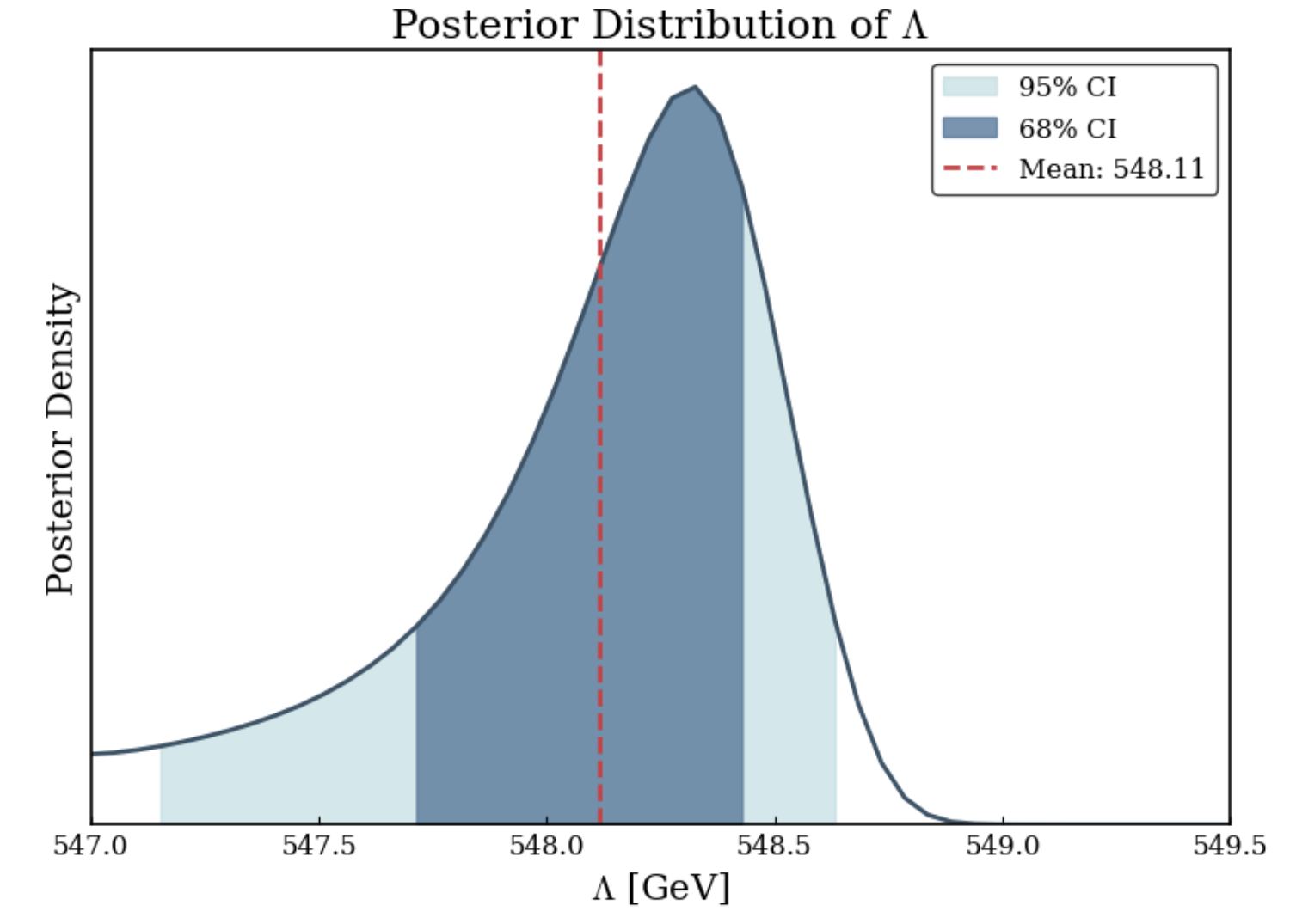}
	\caption{The posterior probability distribution for the reconstruction of the parameter $\Lambda$ for $\mathrm{BP_1}$, based on a simulated analysis for the LISA detector.  The mean of the posterior is indicated by the red dashed line at $\Lambda=548.11~\mathrm{GeV}$ with a standard deviation of $0.37~\mathrm{GeV}$. The shaded regions represent the 68\% (darker blue) and 95\% (lighter blue) CIs.}
	\label{fig:lisalam}
\end{figure}
For LISA, the posterior analysis yields $\Lambda = 548.11 \pm 0.37~\mathrm{GeV}$ for $\mathrm{BP_1}$, with $68\%$ and $95\%$ CIs of $[547.71, 548.43]~\mathrm{GeV}$  and $[547.15, 548.63]~\mathrm{GeV}$, respectively. The injected value $\Lambda_{\rm true} = 548.31~\mathrm{GeV}$ is within the $68\%$ CI of the reconstruction, demonstrating consistency between the reconstructed and injected parameters. This measurement achieves a fractional precision at the sub-percent level.

Compared to TianQin's reconstruction ($\Lambda = 548.12 \pm 0.36~\mathrm{GeV}$), LISA provides comparable constraints. Both detectors demonstrate robust parameter reconstruction capabilities for the dimension-six model when sufficiently strong SGWB signals are present.

\begin{table}[htbp]
\begin{center}
\caption{Comparison between true and reconstructed values of $\Lambda$ for LISA and TianQin detectors. Reconstructed values are shown as posterior mean $\pm$ standard deviation (Std) with 68\% and 95\%  CIs.}
\begin{tabular}{cccccc}
\hline\hline
Detector &   & $\Lambda_{\text{true}}$ [GeV] & $\Lambda$ (Mean $\pm$ Std) [GeV] & 68\% CI [GeV] & 95\% CI [GeV] \\
\hline\hline
\textbf{LISA}    & $\mathrm{BP_1}$   & 548.31 & $548.11 \pm 0.37$ & [547.71, 548.43] & [547.15, 548.63] \\
                 & $\mathrm{BP_2}$   & 549.02 & $549.12 \pm 0.21$ & [548.94, 549.30] & [548.74, 549.45] \\
                 & $\mathrm{BP_3}$   & 550.16 & $554.29 \pm 2.53$ & [551.18, 556.82] & [549.71, 558.53] \\
\hline\hline
\textbf{TianQin} & $\mathrm{BP_1}$   & 548.31 & $548.12 \pm 0.36$ & [547.74, 548.41] & [547.12, 548.65] \\
\hline\hline
\end{tabular}
\label{tab:pa}
\end{center}
\end{table}
We obtained 32 training points by performing first-principles calculations of the effective potential at $\Lambda$ (uniformly spaced in the range $548.0$--$560.0$~GeV, with a spacing of $\approx 0.4$~GeV). All 32 training points are thus obtained from full effective potential computations, so that the machine-learning ensemble is trained exclusively on physically computed data. Tab.~\ref{tab:pa} summarizes the reconstruction results for the model parameter $\Lambda$ across three benchmark scenarios obtained using LISA and TianQin. For each benchmark point, the Tab.~\ref{tab:pa} lists the injected true value $\Lambda_{\text{true}}$, the reconstructed mean and standard deviation, as well as the 68\% and 95\% CIs. A systematic trend in reconstruction precision is evident across the benchmark points. For $\mathrm{BP_1}$ ($\Lambda_{\text{true}} = 548.31~\mathrm{GeV}$), the reconstructed value is $548.11 \pm 0.37~\mathrm{GeV}$, with the true value falling entirely within the 68\% CI, demonstrating excellent reconstruction accuracy. Similarly, $\mathrm{BP_2}$ ($\Lambda_{\text{true}} = 549.02~\mathrm{GeV}$) yields a tight constraint with a standard deviation of $0.21~\mathrm{GeV}$. However, for $\mathrm{BP_3}$, the reconstruction precision deteriorates significantly. The standard deviation increases to $2.53~\mathrm{GeV}$, and the reconstructed mean of $554.29~\mathrm{GeV}$ deviates from the true value ($550.16~\mathrm{GeV}$) by approximately $4.13~\mathrm{GeV}$, although the true value remains within the 95\% CI. This decline in reconstruction precision directly reflects the weakening of the GW signal as $\Lambda$ increases. As $\Lambda$ grows, the strength of the phase transition and consequently the amplitude of the resulting GW signal diminishes, leading to a lower SNR and increased uncertainty in parameter estimation. For $\mathrm{BP_1}$ and $\mathrm{BP_2}$, the signal peak lies well above the LISA foreground noise, allowing high-precision reconstruction. In contrast, the signal from $\mathrm{BP_3}$ approaches or falls below the foreground level, significantly degrading the constraint capability. Nevertheless, even for this weaker signal, LISA is still able to constrain $\Lambda$ with a precision better than $1\%$, demonstrating its robustness in probing marginal GW signals from cosmological phase transitions.

\subsection{Total uncertainty}\label{sec:total}
\subsubsection{Sensitivity to the bubble wall velocity}

\begin{table}[htbp]
\centering
\caption{Sensitivity of the geometric parameter Fisher forecast to variations in the bubble wall velocity, scanned over the deflagration and hybrid regimes.}
\label{tab:summary}
\begin{tabular}{lccccccccc}
\hline\hline
$v_w$ & 0.46 & 0.49 & 0.51 & 0.54 & 0.57~ & 0.59 & 0.62 & 0.65 & 0.68 \\
\hline
$\frac{\Delta\Omega_2}{\Omega_2}$ [\%] & $1.4 \times 10^4$ & $3.1 \times 10^3$ & $4.7 \times 10^2$ & $1.2 \times 10^2$ & 6.3~ & 6.8 & $1.0 \times 10^2$ & $5.2 \times 10^2$ & $3.9 \times 10^3$ \\
\hline\hline
\end{tabular}
\end{table}
We scanned $v_w$ in the range $[0.46,\, 0.68]$, which covers both the deflagration regime and the hybrid regime. For each value of $v_w$, we recomputed the geometric parameters and re-evaluated the Fisher-matrix forecast of the geometric parameters. The results are presented in Tab.~\ref{tab:summary}. At our fiducial choice $v_w=0.57$ for $\mathrm{BP_1}$, TianQin achieves a relative uncertainty on the signal amplitude of approximately 29\%, and the posterior on the model parameter yields $\Lambda=548.12 \pm 0.36~ \mathrm{GeV}$, corresponding to sub-percent statistical precision.

Within the window 
$0.51 \lesssim v_w \lesssim 0.65$ around our fiducial choice, the 
relative Fisher uncertainty on $\Omega_2$ increases by one to two 
orders of magnitude, while outside this window it deteriorates further 
by three to four orders of magnitude. The reconstruction precision on 
$\Lambda$ inherits this sensitivity through the inverse mapping; an 
order-of-magnitude propagation of the Fisher uncertainties suggests 
that the sub-percent precision on $\Lambda$ is broadly preserved 
within the window but would no longer hold outside it. A first-principles determination of $v_w$ from the fundamental parameters of the dimension-six model is important for future work.

\subsubsection{Template uncertainty}
Our reconstruction pipeline assumes that the phase transition is dominated by the sound wave contribution. For strong supercooled transitions, however, MHD turbulence can also contribute to the spectrum and may shift the 
inferred geometric parameters. To quantify this, we computed the combined sound wave and MHD 
turbulence spectrum for $\mathrm{BP_1}$ following the standard parametrization 
of Ref.~\cite{Caprini:2024hue}, with $\kappa_{\rm turb} = \epsilon\,K$ 
and $\epsilon = 0.1$. We find that, at the second break frequency $f_2$, the 
turbulence contribution amounts to a relative shift of $\delta\Omega_2/\Omega_2 
\approx 3\%$. Propagating this shift through the trained machine-learning pipeline yields an 
 offset of $\delta\Lambda_{\rm turb} \lesssim 0.01$~GeV for 
$\mathrm{BP_1}$, which is well below the combined statistical and machine-learning 
uncertainty reported in Tab.~\ref{tab:pa_combined}. The 
bubble collision contribution is negligible in the non-runaway regime relevant 
here. This 
quantitatively justifies restricting our template to the sound wave 
contribution.

\subsubsection{Machine-learning inference uncertainty}

\begin{table}[htbp]
\begin{center}
\caption{Summary of the LOO cross-validation performed on all 32 training points from effective potential calculations.} 
\begin{tabular}{lc}
\hline\hline
Metric & Value \\
\hline
Number of training points              & 32  \\
Data augmentation                       & None \\
RMS reconstruction error ($\sigma_{\mathrm{ML}}$) & 0.11 GeV \\
Maximum absolute error                  & 0.45 GeV \\
\hline\hline
\end{tabular}
\label{tab:loo}
\end{center}
\end{table}
Our pipeline employs an ensemble of four structurally different methods: Gaussian Process Regression, Random Forest, Gradient Boosting Trees, and Multi-Layer Perceptron---whose predictions are combined with performance-based weights. Because these models carry very different inductive biases, systematic overfitting by any single model is penalized in the ensemble average. To provide a direct quantitative test, we have performed a leave-one-out (LOO) cross-validation~\cite{edward2006rasmussen} over all 32 training points in Tab.~\ref{tab:loo}. These 32 points are obtained directly from first-principles effective potential calculations of the dimension-six model (no spline-augmented or otherwise synthetic samples are used in this test).  LOO is a standard diagnostic for assessing generalization error in small-sample regimes.  In the LOO test, each of the 32 training points is held out once, the machine-learning pipeline is retrained on the remaining 31 points, and the held-out value of $\Lambda$ is reconstructed. The resulting root mean square (RMS) error is $\sigma_{\rm ML} = 0.11$~GeV, with a maximum single-point error of $0.45$~GeV. These results demonstrate that overfitting is not a significant concern in our pipeline.

\begin{table}[htbp]
\centering
\caption{Reconstructed $\Lambda$ with combined statistical (stat) and machine-learning (ML) uncertainty.} 
\label{tab:pa_combined}
\begin{tabular}{cccc}
\hline\hline
Detector && $\sigma(\Lambda)^{\rm stat}$ [GeV]  & $\sigma(\Lambda)^{\rm stat+ML}$ [GeV] \\
\hline
\textbf{LISA}    & $\mathrm{BP_1}$ & $0.37$ & $0.39$ \\
                 & $\mathrm{BP_2}$ & $0.21$ & $0.24$ \\
                 & $\mathrm{BP_3}$ & $2.53$ & $2.53$ \\
\hline
\textbf{TianQin} & $\mathrm{BP_1}$ & $0.36$ & $0.38$ \\
\hline\hline
\end{tabular}
\end{table}
Table~\ref{tab:pa_combined} reports the reconstructed $\Lambda$ together with combined statistical and machine-learning uncertainties for each benchmark point. The combined uncertainty is obtained by adding the \texttt{PolyChord} posterior width and the LOO RMS error in quadrature
\begin{equation}
  \sigma_\Lambda^{\rm stat+ML}
    = \sqrt{\sigma_{\rm stat}^2 + \sigma_{\rm ML}^2}
    .
\end{equation}
Theoretical systematic uncertainties arising from the effective potential calculation are discussed separately above and are not included in this estimate.  We note that, including the machine-learning inference error increases the total statistical uncertainty by at most a few percent relative to the purely statistical estimate, so the sub-percent-level statistical precision is preserved.

\section{Conclusions and discussions}\label{sec:sum}

We have investigated the capability of the space-based GW observatories TianQin and LISA to reconstruct both the spectral and particle-physics parameters of EWPT signals within the dimension-six SMEFT framework. This model introduces a cutoff scale $\Lambda$ as the only effective model parameter and provides a well-motivated scenario capable of realizing a SFOPT, leading to an observable SGWB. We have implemented a comprehensive reconstruction pipeline that employs Fisher-matrix analysis and Bayesian nested sampling via \texttt{PolyChord} to reconstruct the geometric parameters, incorporating TDI channel noise and astrophysical foregrounds, then utilizes machine-learning techniques to map these geometric observables to the model parameter $\Lambda$. Using the SW spectrum based on the DBPL template, this approach directly links GW observations to the fundamental theoretical framework.

We have evaluated the parameter reconstruction performance of TianQin and LISA using three representative benchmark scenarios ($\mathrm{BP_1}$--$\mathrm{BP_3}$) of the dimension-six SMEFT model. For $\mathrm{BP_1}$, which has a break frequency of approximately $1.10~\mathrm{mHz}$, TianQin achieves an amplitude reconstruction precision of about 29.19$\%$, with a $1\sigma$ uncertainty on the reconstruction of model parameter of $\sigma_\Lambda^{\rm stat+ML} = 0.38~\mathrm{GeV}$, corresponding to sub-percent statistical precision.
However, TianQin's reconstruction capability is limited, and the signals corresponding to the other benchmarks ($\mathrm{BP_2}$, $\mathrm{BP_3}$) fall below its sensitivity threshold and cannot be reconstructed.
In contrast, LISA, benefiting from a broader frequency coverage, successfully reconstructs parameters across all three benchmark scenarios.
For $\mathrm{BP_1}$ (break frequency $28.07~\mu\mathrm{Hz}$), LISA achieves a relative uncertainty of 30.15$\%$ in the signal amplitude and $\sigma_\Lambda^{\rm stat+ML} = 0.39~\mathrm{GeV}$.
For $\mathrm{BP_2}$ (break frequency $273.62~\mu\mathrm{Hz}$), the amplitude precision improves to 20.69$\%$, with $\sigma_\Lambda^{\rm stat+ML} = 0.24~\mathrm{GeV}$.
For sufficiently strong signals, LISA achieves sub-percent precision in $\Lambda$ comparable to TianQin.
Even for weak signals near the detection threshold ($\mathrm{BP_3}$), LISA maintains better than 1$\%$ statistical accuracy in reconstructing the model parameter $\Lambda$.
Although LISA demonstrates stronger overall reconstruction performance for the dimension-six model, the complementarity between TianQin and LISA remains important.
For signals of comparable amplitude, each detector provides the tightest constraints when the signal frequency lies within its optimal sensitivity band.
Nevertheless, both detectors share the same fundamental limitation: reconstruction precision degrades dramatically when the signal amplitude approaches or falls below the level of the astrophysical foreground, irrespective of frequency.
This finding emphasizes that multi-detector networks---while advantageous for probing a variety of EWPT scenarios---cannot fundamentally overcome the foreground-limited regime that constrains weak signals. 

It is important to note that these conclusions are based on several assumptions. The sub-percent precision quoted here should be understood under the following conditions: (i) it refers to the statistical reconstruction capability of the detectors and incorporates the machine-learning inference uncertainty; (ii) it is established at a fixed bubble wall velocity, and we have verified that the sub-percent precision is preserved within a narrow window but degrades by orders of magnitude outside this window; (iii) it does not include theoretical uncertainties from the 4d perturbative treatment of the finite-temperature effective potential---including resummation scheme dependence, gauge dependence, and renormalization scale dependence---which can introduce $\mathcal{O}(1)$ or larger uncertainties in the predicted GW amplitude. Consequently, our sub-percent precision results should be interpreted as the statistical reconstruction capability of the detectors under idealized theoretical assumptions, representing the best achievable precision in this idealized limit.

In addition, we used matched templates for both signal injection and reconstruction, thereby neglecting possible systematic errors arising from theoretical uncertainties in phase transition dynamics and numerical simulations.  Consequently, our results should be regarded as an approximate upper bound on the achievable reconstruction precision under idealized conditions. Robust parameter inference from real LISA/TianQin data will require: 
(i) refined theoretical modeling of SGWB production mechanisms with quantified uncertainties, (ii) improved foreground characterization and subtraction techniques, and (iii) data analysis pipelines validated through realistic data challenges.
Although some quantitative assessments presented here are benchmark-specific, they nonetheless provide representative guidance regarding the dominant sensitivities and systematic limitations of the reconstruction framework.
Joint observations with TianQin and LISA would combine their complementary sensitivities and maximize the achievable constraints on model parameters. Moreover, the combination of GW detections with particle physics experiments, particularly future colliders probing the $\mathrm{TeV}$ scale, offers the prospect of multi-messenger verification of EWPT physics. We can conclusively test BSM scenarios and uncover the fundamental mechanisms in the early universe only through such coordinated strategies.

\begin{acknowledgments}
We are grateful to Siyu Jiang, Dayun Qiu, Quan Chen for valuable advices on this manuscript. We thank Philipp Schicho for valuable discussions on the theoretical uncertainties of the effective potential.
This work is supported by the National Natural Science Foundation of China (NNSFC) under Grant No.12475111, No.12205387, and the Fundamental Research Funds for the Central Universities, Sun Yat-sen University.
\end{acknowledgments}

\bibliographystyle{apsrev}
\bibliography{axionref}
\end{document}